\documentclass[bibyear]{aa}  

\usepackage{graphicx}
\usepackage{natbib}
\usepackage{txfonts}
\usepackage{epstopdf}
\usepackage{multirow}
\usepackage{url}
\usepackage{textcomp}
\newcommand{\vpeg}{V374\,Peg}
\newcommand{\HA}{H$\alpha$}
\newcommand{\kms}{km\,s$^{-1}$}
\newcommand{\msun}{$M_\odot$}
\newcommand{\rsun}{$R_\odot$}
\begin{document}

\title{Investigating magnetic activity in very stable stellar magnetic fields}
\subtitle{Long-term photometric and spectroscopic study of the fully convective M4 dwarf V374 Peg }

   \author{K. Vida\inst{1}
          \and
          L. Kriskovics\inst{1}
		  \and
		  K. Ol\'ah\inst{1}
		  \and
		  M. Leitzinger\inst{2}
		  \and
		  P. Odert\inst{3,2}
		  \and
  		  Zs. K\H{o}v\'ari\inst{1}
		  \and
		  H. Korhonen\inst{4,5}
		  \and
		  R. Greimel\inst{2}
		  \and 
		  R. Robb\inst{6}
		  \and
		  B. Cs\'ak\inst{7}
		  \and
		  J. Kov\'acs\inst{7}
		  }

   \institute{
			Konkoly Observatory, MTA CSFK, H-1121 Budapest, Konkoly Thege M. út 15-17, Hungary\\ 
			\email{vidakris@konkoly.hu}
			\and
			University of Graz, Institute of Physics, Department for Geophysics, Astrophysics, and Meteorology, NAWI Graz, Universitätsplatz 5, A-8010 Graz, Austria
			\and
			Space Research Institute, Austrain Academy of Sciences, Schmiedlstrasse 6, A-8042 Graz, Austria
			\and
			Finnish Centre for Astronomy with ESO, University of Turku, V\"ais\"al\"antie 20, FI-21500 Piikki\"o, Finland
			\and
			Dark Cosmology Centre, Niels Bohr Institute, Copenhagen University, Juliane Maries Vej 30, 2100, Copenhagen {\O}, Denmark
			\and
			University of Victoria and Guest Observer, Dominion Astrophysical Observatory, Canada 
			\and 
             ELTE Gothard Astrophysical Observatory, H-9704 Szombathely, Szent Imre herceg \'ut 112, Hungary}

   \date{Received September 15, 1996; accepted March 16, 1997}

\abstract{
The ultrafast-rotating ($P_\mathrm{rot}\approx0.44\,d$) fully convective single M4 dwarf V374\,Peg is a well-known laboratory for studying intense stellar activity in a stable magnetic topology. As an observable proxy for the stellar magnetic field, we study the stability of the light curve, and thus the spot configuration. We also measure the occurrence rate of flares and coronal mass ejections (CMEs).
We analyse spectroscopic observations, $BV(RI)_C$ photometry covering 5 years, and additional $R_C$ photometry that expands the temporal base over 16 years.
The light curve suggests an almost rigid-body rotation, and a spot configuration that is stable over about 16 years, confirming the previous indications of a very stable magnetic field. We observed small changes on a nightly timescale, and frequent flaring, including a possible sympathetic flare. The strongest flares seem to be more concentrated around the phase where the light curve indicates a smaller active region. Spectral data suggest a complex CME with falling-back and re-ejected material, with a maximal projected velocity of $\approx 675$\,\kms . We observed a CME rate much lower than expected from 
 extrapolations of the solar flare--CME relation to active stars.
} 


   \maketitle
%

\section{Introduction}    

\vpeg{} is an M4 dwarf \citep{1995AJ....110.1838R} 
that first raised interest when \cite{Greimel:1998:flares} detected frequent strong flares on the star. \cite{Batyrshinova:2001:flare} later observed several flares, including a superflare event, having 11$^m$ amplitude in $U$ filter. \cite{Montes:2001:castorgroup} found, that \vpeg{} is part of the Castor kinematic group, thus has an age of approximately 200\,Myr. This young age is also supported by \cite{Vidotto:2011:wind}, who found evidence for a strong coronal wind.

Due to its high activity level and relative brightness ($V\approx11.\!\!^{\rm m}5$), \vpeg{} is a popular astrophysical laboratory for studying stellar magnetic fields in young fully convective stars.
\cite{Donati:2006:zdi} studied the magnetic field of \vpeg{} using Zeeman--Doppler imaging on data covering three non-consecutive rotations. They found that the star was rotating as a solid body, and a dipole-like axisymmetric magnetic field. This result challenged theoretical dynamo models, which predicted either rigid-body rotation with non-axisymmetric magnetic field (e.g. \citealt{1999A&A...346..922K}), or axisymmetric fields, but with differential rotation (e.g. \citealt{2006ApJ...638..336D}). 
 A recent study by \cite{2015ApJ...813L..31Y} proposed a dynamo model that can explain solid-body rotation with dipole-like field, and predicted long-term stable magnetic configuration.
Based on observations from three epochs spanning over a year \cite{Morin:2008:zdi} detected a very weak differential rotation and concluded that the magnetic field of \vpeg{} is stable on this time scale.

In this paper we present analysis of photometric and spectroscopic data spanning over more than a half decade.

\section{Observations}

\begin{figure}
	\centering
	\includegraphics[width=0.5\textwidth]{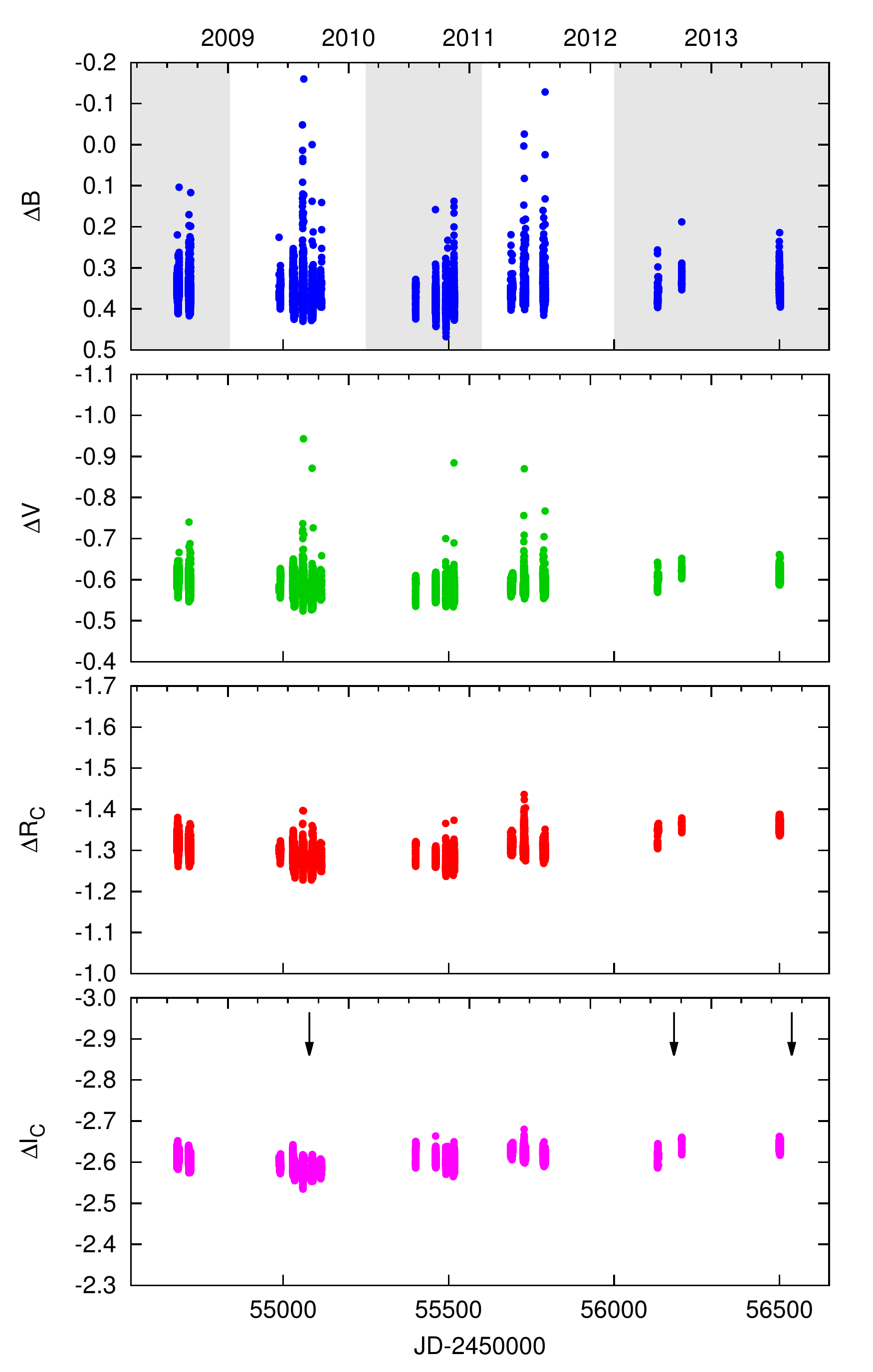}
	\caption{$BV(RI)_C$ photometry of V374\,Peg from the Piszk\'estet\H{o} observations. Ranges marked in the $\Delta B$ light curve were used for flare statistics (see Section \ref{sect:flarestat}). Arrows over the $\Delta I$ light curve mark the dates of the spectroscopic observations during this time.}
	\label{fig:lcall}
\end{figure}

\begin{figure}
	\centering
	\includegraphics[width=0.5\textwidth]{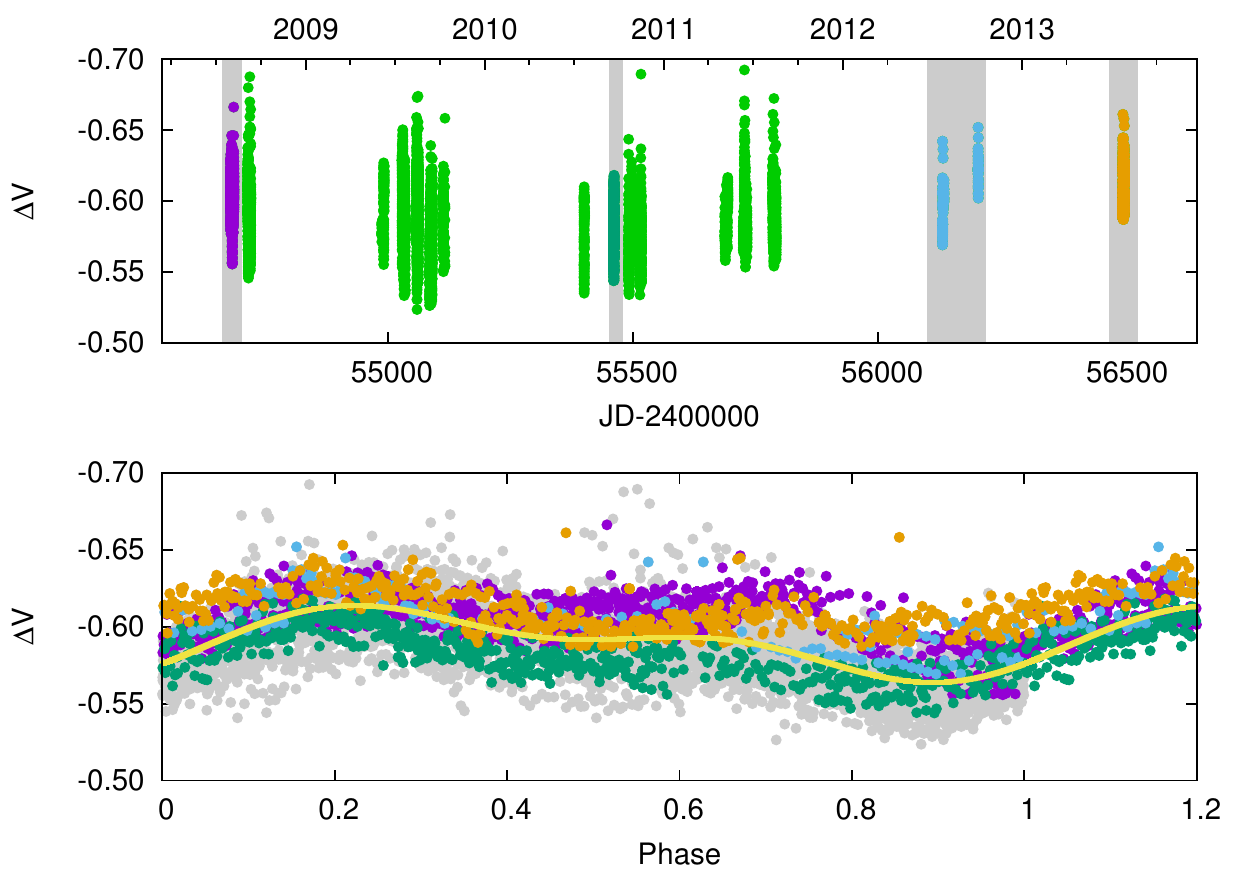}
	\caption{$\Delta V$ light curve of \vpeg{} (top), and the marked parts phased at the bottom. Continuous line shows a model fit with two spots (see Sect. \ref{sect:sml}). Note, that the plot does not show most of the flares.}
	\label{fig:phasedlc}
\end{figure}

Photometric observations in $BV(RI)_C$ passbands were carried out using the 1-m RCC telescope of Konkoly Observatory at Piszk\'estet\H{o} Mountain Station equipped with a Princeton Instruments $1300\times1300$ CCD. The observations cover 71 nights in total between 2008 July 31 and 2013 July 28 (HJDs 2454679--2456502). 
Data reduction was done using standard IRAF\footnote{IRAF is distributed by the National Optical Astronomy Observatory,
which is operated by the Association of Universities for Research in Astronomy, Inc., under cooperative agreement with the National Science
Foundation.} 
procedures. Differential aperture photometry was done using the DAOPHOT package. The resulting $BV(RI)_C$ light curves are shown in Fig. \ref{fig:lcall}. 
All $\Delta V$ data of this dataset is plotted together against the rotational phase (see Sect. \ref{sect:fourier}) in Fig. \ref{fig:phasedlc}. We used GSC 02215-01602, as comparison star for our light curves.

Additional CCD photometric data in Cousins $R$ filter were obtained from the automated 0.5-m telescope  of the Climenhaga Observatory of the University of Victoria from 1998 and 2000, that were partly published in \cite{Greimel:1998:flares}.

\begin{table}
	\caption{Summary of spectroscopic observations}
	\label{tab:specobslog}
	\centering
	\begin{tabular}{cccc}
	\hline
	Date&HJD&Telescope&\# of spectra\\
	\hline
	\hline
2005/08/18--08/23 &2453601--06& CFHT & 245 \\
2006/08/04--08/12 &2453952--60& CFHT & 88  \\
2009/09/03--09/06 &2455078--81& CFHT & 96 \\
2012/09/08--09/12 &2456179--82 & RCC &13\\
2013/08/12--09/20 &2456517--55 & RCC &11\\
\hline

	\end{tabular}
\end{table}

The spectroscopic observations in 2012/2013 were carried out by the 1-m RCC telescope, using the eShel spectrograph of
the Gothard Astronomical Observatory, Szombathely, Hungary, in the spectral range 4200--8700\AA{}  with a resolution of $R \approx 11\,000$ \citep{gaospec}. The wavelength calibration was done using a ThAr lamp. Reduction of these data was carried out using standard IRAF procedures.

Additional spectra were downloaded from the public database of Canada--France--Hawaii Telescope (CFHT)\footnote{
Available at \url{http://www.cadc-ccda.hia-iha.nrc-cnrc.gc.ca/en/cfht/}}. 
The spectra were obtained between 2005 and 2009 by ESPaDOnS \citep{espadons} with a spectral resolution of $R\approx65\,000$ covering the range between 3700--10400\,\AA . The spectroscopic observations are summarised in Table \ref{tab:specobslog}.

\section{Analysis}    

\subsection{Mass}
\label{sect:mass}
Using the empirical absolute magnitude--mass calibration for very low mass stars ($M<0.6$\msun) of \cite{Delfosse:2000:mass}, we can get an empirical mass value following \cite{Morin:2008:zdi}. According to the The Two Micron All Sky Survey (2MASS), the magnitude of \vpeg{} in $J$, $H$ and $K$ passbands are 7.635, 7.035, 6.777 magnitudes \citep{2massb}, which translate to the absolute magnitudes of 7.887,  7.287 and  7.029 respectively using the parallax of 112.33\,mas from the Hipparcos survey \citep{hipparcos}, if we neglect interstellar absorption (as the distance of the object is only 8.9\,pc). These magnitudes yield to mass values of 0.282, 0.285 and 0.281$M_\odot$ for the three passbands, respectively.

Using the magnitude--mass and magnitude--radius calibrations for M-dwarfs of \cite{2015ApJ...804...64M} based on the $K$ magnitudes, we get a somewhat different result as previously: 0.338\,\msun{} for the mass and 0.335\,\rsun{} for the radius, in good agreement with \cite{2008MNRAS.390..567M}.
We can  conclude that from these different methods the mass of \vpeg{} is $0.30\pm0.03$\msun  (the error estimated using the standard deviation of the different mass values).

\subsection{Spectral energy distribution: is there an IR excess?}
\begin{figure}
	\centering
	\includegraphics[width=0.5\textwidth]{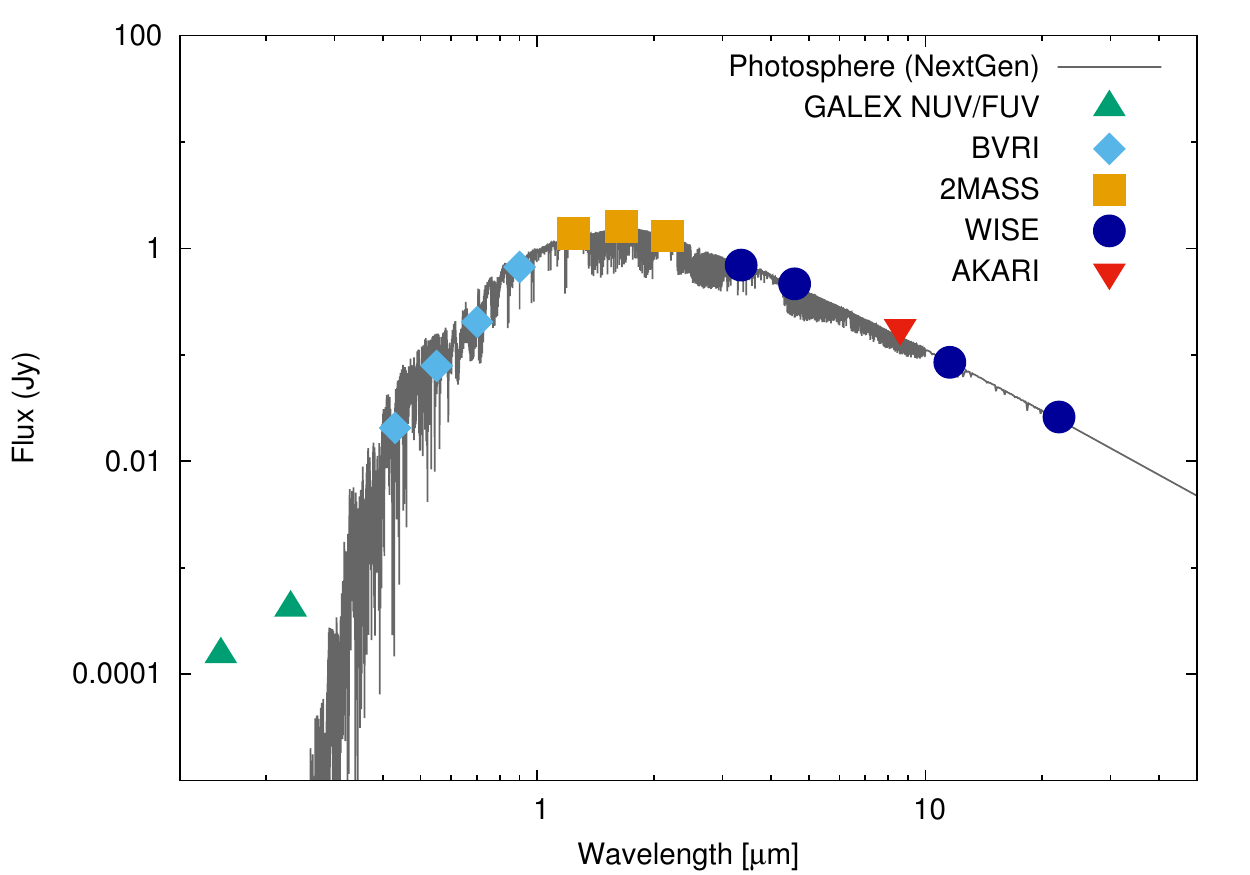}
	\caption{Spectral energy distribution of \vpeg{} from the ultraviolet to mid-infrared using NextGen model atmospheres (see text for more details). Errors bars are smaller than the symbols used for the plot.}
	\label{fig:sed}
\end{figure}
\cite{irexcess} studied active RS CVn systems, and found  infrared excess in 5 of 12 cases, which the authors interpreted as evidence for an absorbing shell, a result of strong stellar wind. 
Such excess was (unsuccessfully) searched also in active single stars \citep{2005AJ....130.1231P}, but was found in many fast rotating evolved stars  \citep{2015ApJ...801...54R}. The latter authors also considered the dust being the result of planetary collisions.

If we neglect interstellar reddening, we can easily determine the intrinsic infrared colours of \vpeg : $(J-K)_0=0.858$ and $(H-K)_0=0.258$ using the 2MASS magnitudes \citep{2massb}. Comparing these values to the intrinsic colours of a main-sequence M4 star 
($J-K=0.839$ , $H-K=0.282$) based on the work of \cite{2013ApJS..208....9P}, 
we find a difference of $0.02$ and $-0.02$ magnitudes, which is around the typical error of 0.03 magnitudes of the 2MASS observations. 

The suggestion of \cite{irexcess}, i.e., that stellar winds due to high level of activity can induce infrared excess, was further tested 
by plotting the spectral energy distribution (SED), see Fig. \ref{fig:sed}. We used the \verb+VOspec+ tool\footnote{Available at \url{http://www.sciops.esa.int/index.php?project=SAT&page=vospec}} that is capable of obtaining both multi-wavelength photometric observations and theoretical spectra. We downloaded ultraviolet observations from the GALEX survey \citep{galex}, and infrared observations from the 2MASS \citep{2massb}, WISE \citep{wise} and AKARI \citep{akari} surveys. We also plotted our $BV(RI)_C$ observations from the Piszk\'estet\H{o} RCC telescope.
To describe the photosphere we used the NextGen \citep{nextgen} model with the astrophysical parameters of \vpeg . All the points fit the model photosphere, only the measurements in the ultraviolet regime show a very strong excess consistent with the high chromospheric activity level.
Thus we conclude that \vpeg{} does not have a significant colour excess in the infrared bands, and does not possess a hot dust shell.

\subsection{Fourier analysis}
\label{sect:fourier}
\begin{figure}
	\centering
	\includegraphics[width=0.5\textwidth]{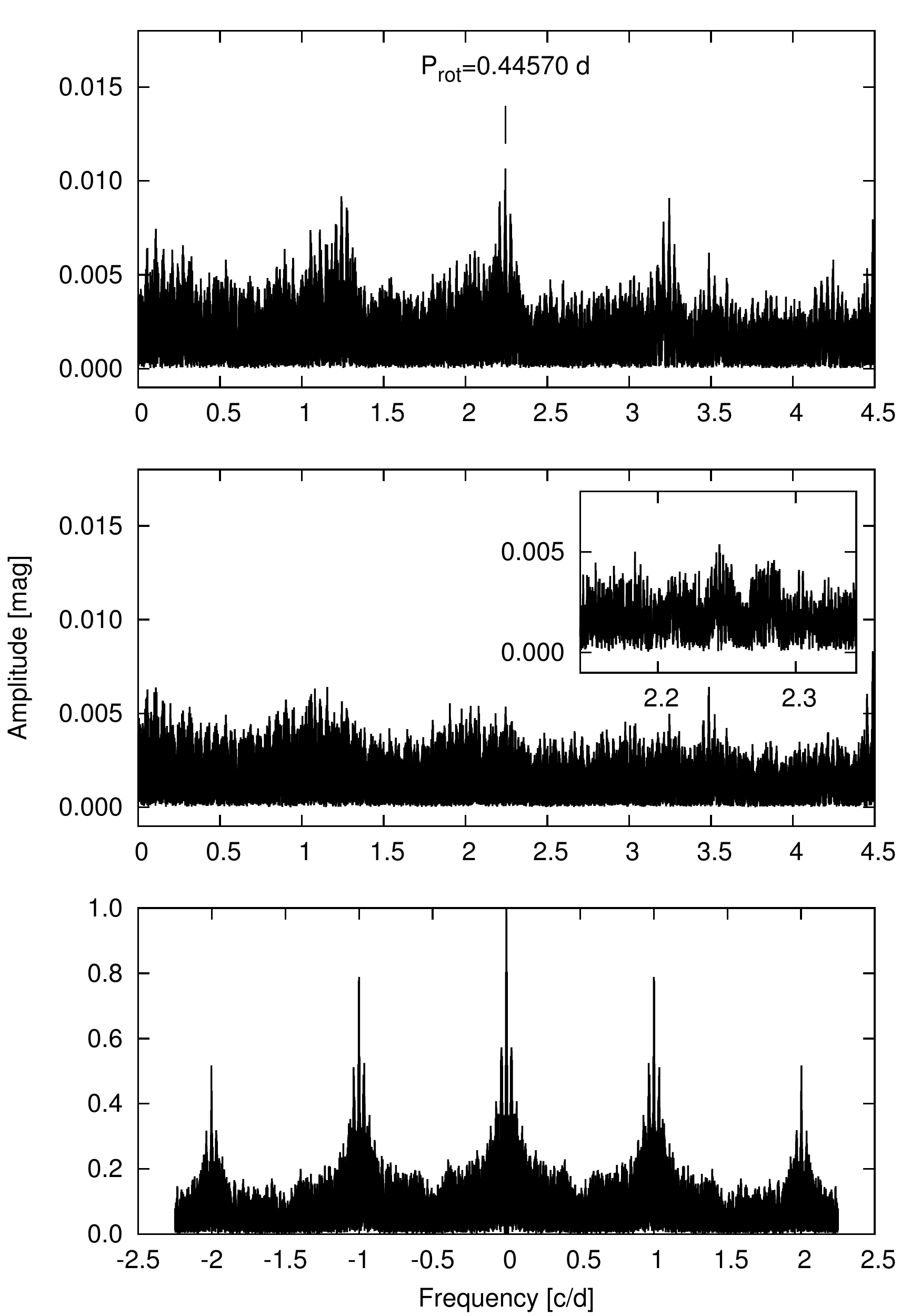}
	\caption{Fourier spectrum of the $\Delta R$ light curve (top), the spectrum pre-whitened with the rotation frequency (middle), the inset showing a zoom-in to the rotation area, and the spectral window (bottom).}
	\label{fig:fourier}
\end{figure}
Fourier analysis of the photometric $\Delta R_C$ data was performed using MUFRAN\footnote{\url{http://www.konkoly.hu/staff/kollath/mufran.html}} \citep{mufran} -- a code that can do discrete Fourier transformation of data and pre-whiten light curves with the detected frequencies -- since this filter has the longest time base of 
16 years (1998--2013), as the observations at the Climenhaga Observatory \citep{Greimel:1998:flares} were done using only this filter. First, due to the different comparison stars used at the two observatories, the 1998 and 2000 datasets were shifted to the mean $R_C$ of the Konkoly dataset, next, small amplitude long-term trends were removed from the data. No systematic long-term change resembling to a spot cycle, or part of it, was found.
The final Fourier spectrum and the spectral window is shown in Figure \ref{fig:fourier}. The peak corresponding to stellar rotation is at 
$P_\mathrm{rot}=0\fd44 57 0\pm0\fd00 00 1$
which is quite close to the finding of 0.445679 days in \cite{2011IAUS..273..460V}. We found the same result, $P_\mathrm{rot}=0.44570$ days using the SLLK method (string length method with Lafler--Kinman statistic, see \citealt{sllk}).
In this paper we used the following ephemeris:
\begin{equation}
\mathrm{HJD} = 2453601.78613 + 0.44 57 0 \times E,
\end{equation}
where HJD$_0=2453601.78613$ is taken from \cite{2011IAUS..273..460V} and $E$ is the cycle count.
After pre-whitening the Fourier spectrum with the signal of the rotation period, 
a very small peak (with $0\!\!.^{\rm m}005$ amplitude) corresponding to $P=0.445518$ days in the Fourier spectrum is seen. If we accept this as a signal originating also from the rotation, this would yield a differential rotation shear of $|\alpha|=|\Delta P/P| \gtrsim 0.0004$, which is very close to a rigid-body rotation, in agreement with the findings of \cite{2013AN....334..972V} and that of \cite{Morin:2008:zdi} ($\alpha=0.0004$) and \cite{Donati:2006:zdi} ($\alpha=0.0014$). 
On GJ~1243, an other M4 star with fast rotation of 0.59 days, \cite{2015ApJ...806..212D} found a similar differential rotation ($|\alpha|=0.0011$) to our estimate on V374 Peg.

\subsection{Light curve modelling}
\label{sect:sml}
Light curve modelling was done using the code {\sc SpotModeL} \citep{sml} that applies an analytic approach, and uses at most three circular starspots of homogeneous temperature to fit the photometric data. Limb darkening coefficients were adopted from \cite{limbdarkening}. From the colour indices we can estimate the spot temperatures, we got about 3250\,K for $B-V$ and $V-I_C$ indices ($3267\pm 5$K and $3244\pm8$K respectively, with formal errors), supposing an unspotted temperature of 3400\,K corresponding to the dM4 spectral type. Thus, from the photometric spot temperature model, we concluded, that the spots are  $\approx$150\,K cooler than the unspotted surface.

During the $\approx$5 years of observations, the light curve was very stable (see Fig. \ref{fig:phasedlc}), it can be described reasonably well by a two-spot model.
From the $\Delta V$ light curve, we can estimate the spot parameters
$\lambda_1=  143 \pm2^\circ$, $\beta_1=67 \pm 3^\circ$, $\gamma_1=42 \pm 3^\circ$ and
$\lambda_2= 342\pm2^\circ$, $\beta_2=71 \pm 4^\circ$, $\gamma_2=36 \pm 4^\circ$, 
where $\lambda$, $\beta$ and $\gamma$ means the spot longitude, latitude and radius, respectively.  According to this model, the cool spots cover approximately 43\% of the stellar surface.
This solution is of course not unique, and the parameters are not independent. Spot latitude is the most uncertain, since photometric observations hold only very little information on the latitude of the spots, and  the (formal) errors do not indicate the uncertainty of the inclination.

\section{Results}     

\subsection{Photometric flares}
\label{sect:flarestat}

\begin{figure}
	\centering
	\includegraphics[width=0.48\textwidth]{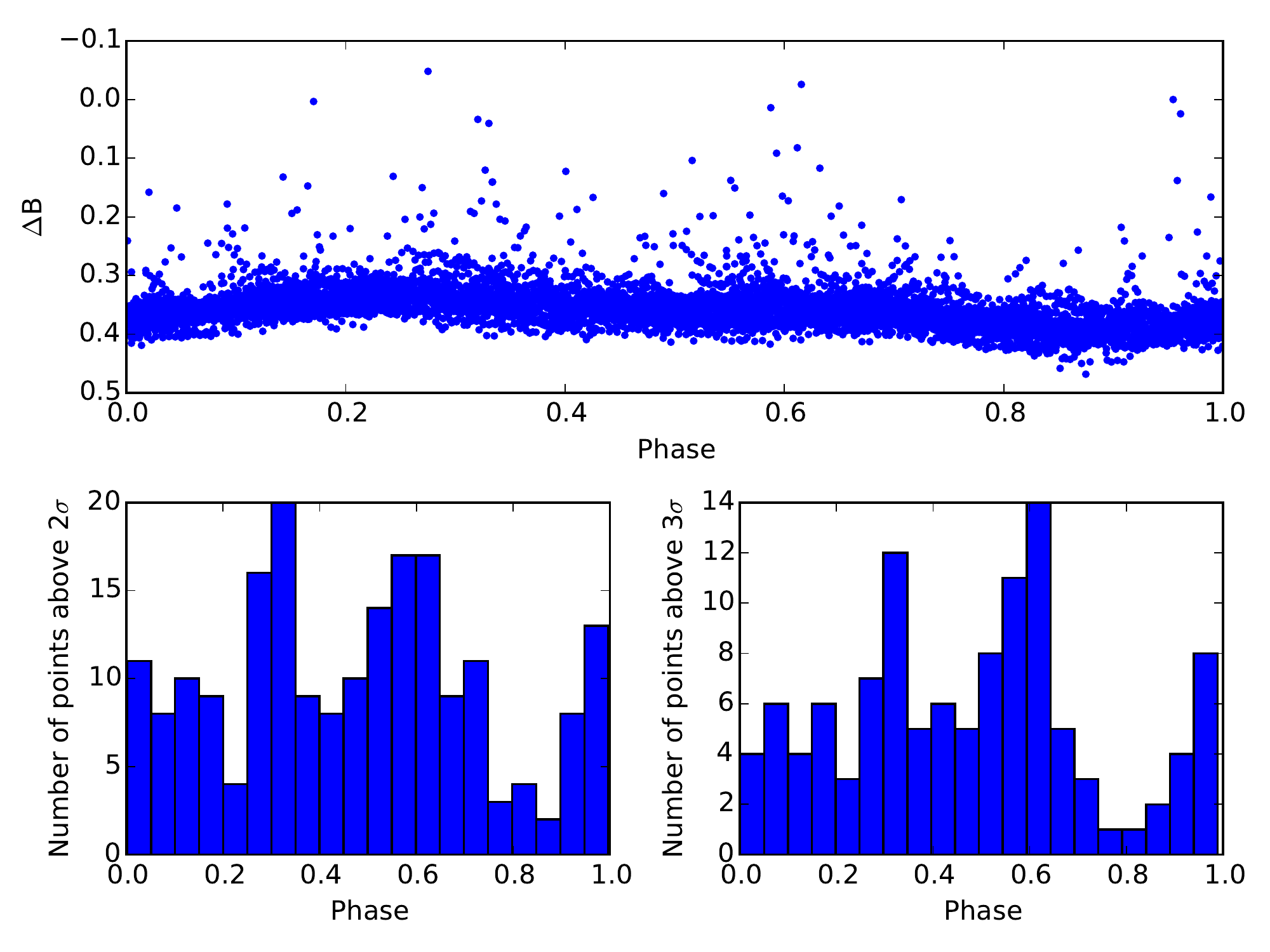}
	\caption{Phased $\Delta B$ light curves as used for flare statistics (top) and number of light curve points above $2\sigma/3\sigma$ after removing the effect of spottedness (bottom).
	Note that the top plot does not show the whole magnitude range of flares to better visualise the spotted light curve.}
	\label{fig:flarehist}
\end{figure}

To study flares on the star, we applied time series spot models to remove the contribution of cool spots \citep[see also][]{v405and}. This was performed by {\sc SpotModeL}, modelling the four-colour light curves that were cleaned from flares, in a time window of 100 days, and shifting this window by 30-day steps, thus reaching a better-fitting analytic solution to the observations. This analytic solution was then subtracted from the original observations, leaving only the flares and small scatter/variations on the level of a few hundredth of magnitude.

All $\Delta B$ observations are plotted in Fig. \ref{fig:flarehist} where the highest amplitude flares are truncated showing  the spotted light curve better. We can see in the top panel it is seen that flares occur in every phase of the rotational modulation. This could mean many different flaring regions even near the pole so the flares are seen regardless of rotational phase. 
Such polar spot, would cause only very small variations in the light curve, so other (non-polar) active regions must be still present. 

Since the time resolution of the data is not too high (around 3 minutes on average) small amplitude and short flares could be represented only by 1--2 deviating data points. The two histograms on the bottom panel of Fig. \ref{fig:flarehist} show the distribution of such deviating points over $2\sigma$ (left) and $3\sigma$ (right) above the folded light curve. A certain increase of such events is seen between phases 0.3--0.7 which suggests a more active centre of flaring on the stellar surface.

\begin{figure}
	\centering
	\includegraphics[width=0.48\textwidth]{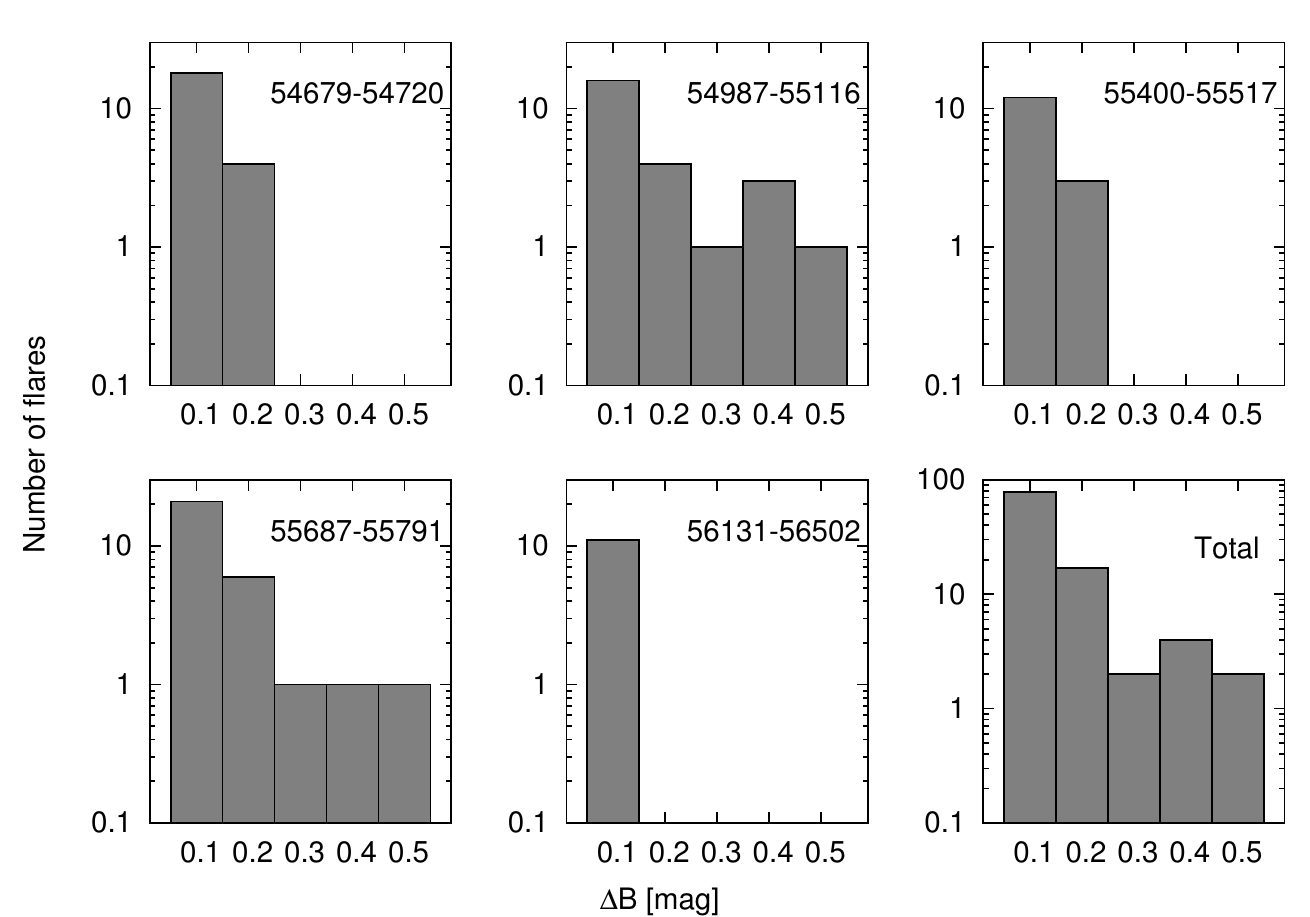}
	\includegraphics[width=0.45\textwidth]{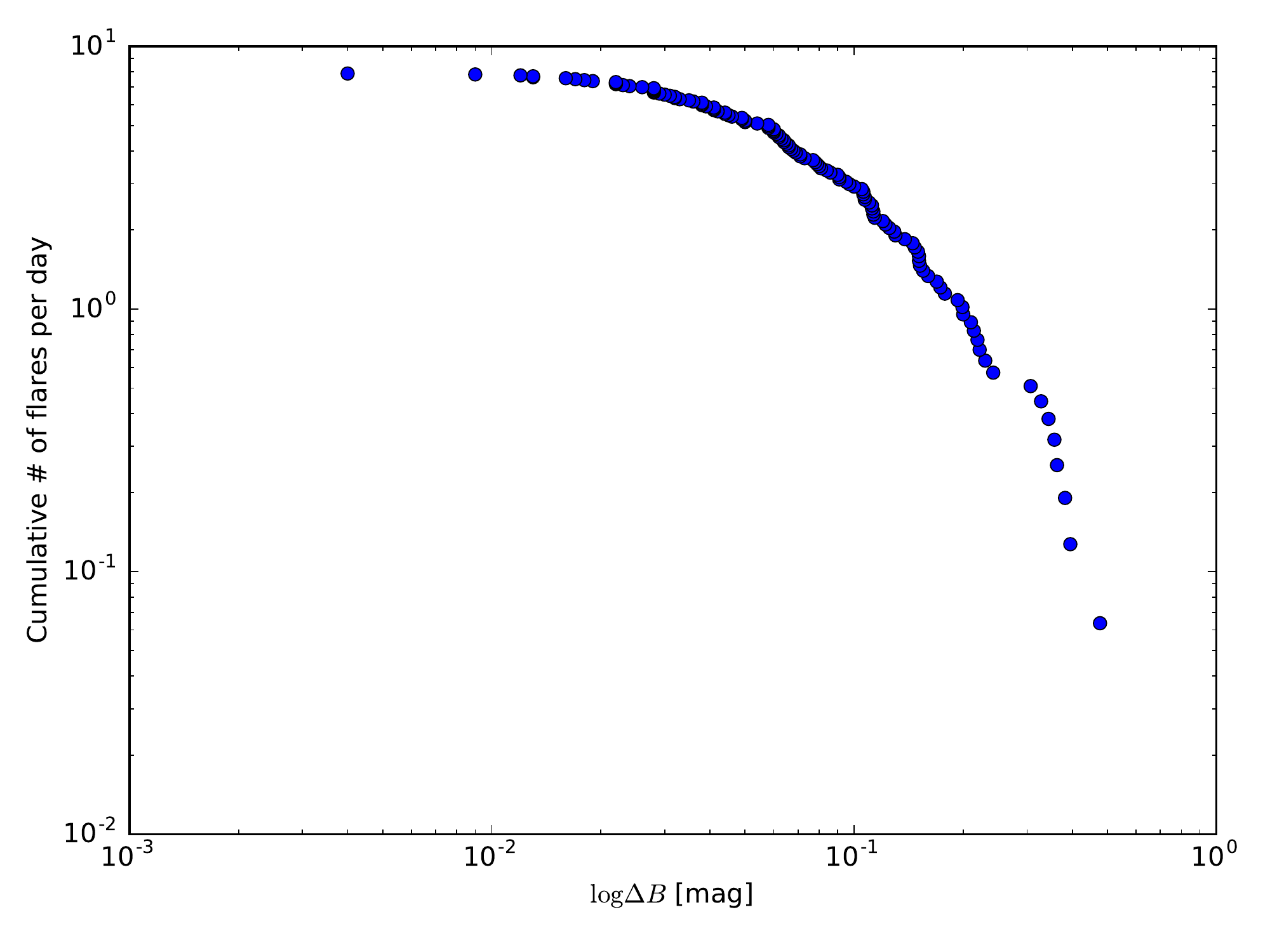}

	\caption{Top plots: numbers of flares are shown to their peak $B$ brightness. Bottom plot: cumulative number of flares per day vs. flare amplitudes.}
	\label{fig:flarestat}
\end{figure}

\begin{table*}
\caption{Summary of flare statistics from the $\Delta B$ light curve.}
\label{tab:flarestat}
\centering
\begin{tabular}{c|c|c|cc|cc}
\hline
\multirow{2}{*}{HJD-2450000} & \# of &Observed& \multicolumn{4}{c}{Flares stronger than}\\
& nights &hours& $0.\!\!^{\rm m}05$ &frequency (h$^{-1}$)& $0.\!\!^{\rm m}25$ &frequency (h$^{-1}$)\\
\hline
\hline
4679--4720 & 13 & 84.0 & 22 & 0.26& 0 & 0.00\\
4987--5116 & 23 & 125.5& 25 & 0.15& 5 & 0.04\\
5400--5517 & 10 & 62.1 & 15 & 0.21& 0 & 0.00\\
5687--5791 & 17 & 71.4 & 30 & 0.42& 3 & 0.04\\
6131--6502 & 8  & 34.5 & 11& 0.32 & 0 & 0.00\\
\hline
\end{tabular}
\end{table*}

For the determination of the flare frequency, we used the method of \cite{flarestat}, who defined the mean occurrence rate of flares as:
\begin{equation}
\nu=\frac{n}{\sum t},
\end{equation}
$n$ being the number of observed flares, and $\sum t$ (hr) the monitoring time. We selected flare events using the unspotted $\Delta B$ light curve (see Sect. \ref{sect:sml}) by visual inspection, selecting flares that were brighter than 0.05 magnitude. To detect possible variations in the flaring activity, the $\Delta B$ light curve was divided into five blocks (see top panel of Fig. \ref{fig:lcall}). The flare frequencies at the different blocks of the observations are summarised in Table \ref{tab:flarestat}. A histogram of flares with different brightness is shown in Fig. \ref{fig:flarestat}. The  occurrence rate seems to be around 0.2--0.3\,h$^{-1}$, relatively stable during the observations. 
However, flares with peak brightness higher than $0.\!\!^{\rm m}25$ were only found in the second and fourth block. In  Fig. \ref{fig:flarestat} we also show the cumulative flare frequency distribution  (log number of flares that have larger amplitude than $\log B$, see e.g. \citealt{FFD1,FFD2}) as function of the flare amplitudes in $B$ passband.

\onlfig{
\begin{figure}
	\centering
	\includegraphics[width=0.5\textwidth]{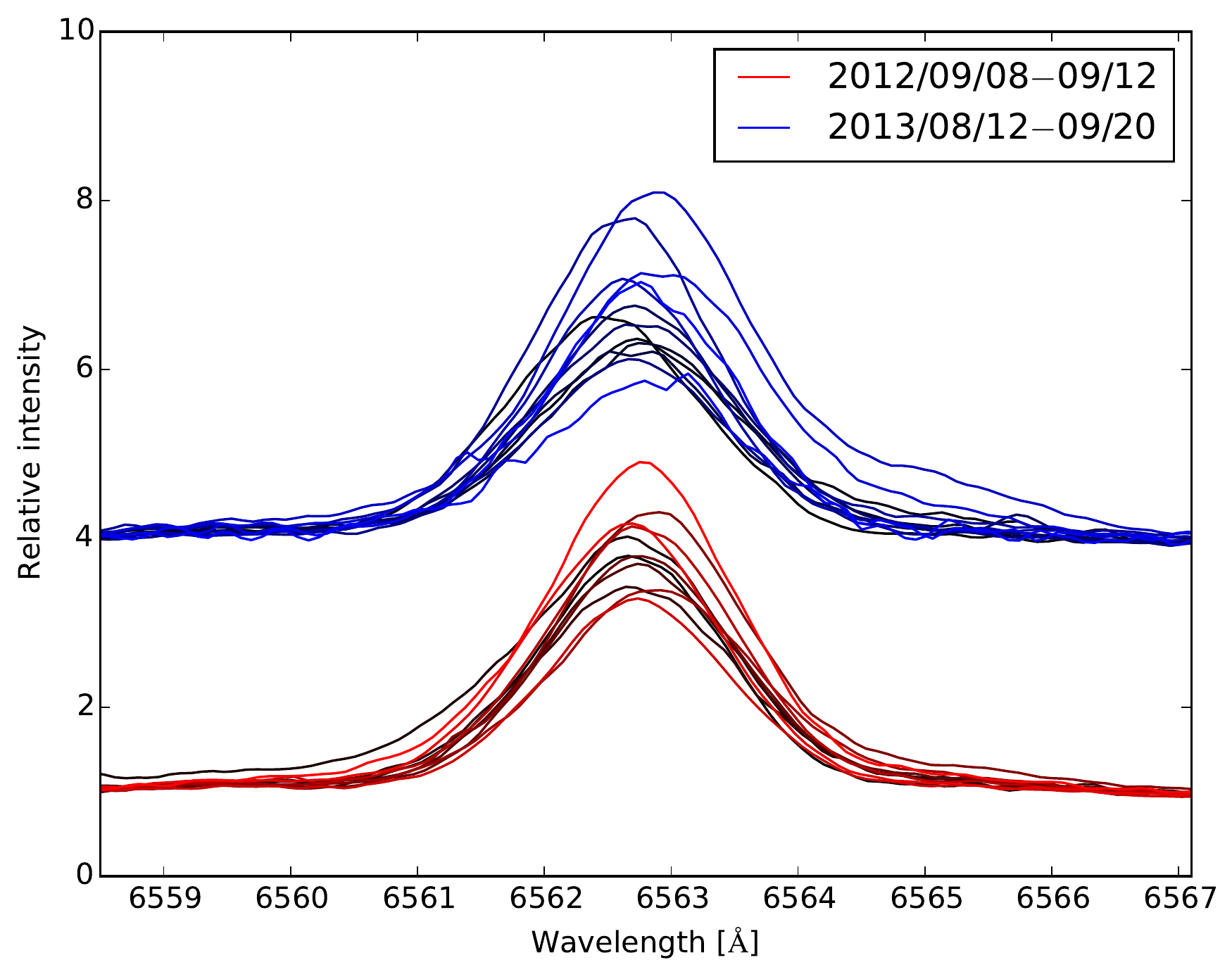}
	\caption{\HA{} profiles from the 1-m RCC telescope (2012/2013), (colour saturation increasing with time).}
	\label{fig:halpha}
\end{figure}
}
\begin{figure*}
	\centering
	\includegraphics[width= 0.32\textwidth]{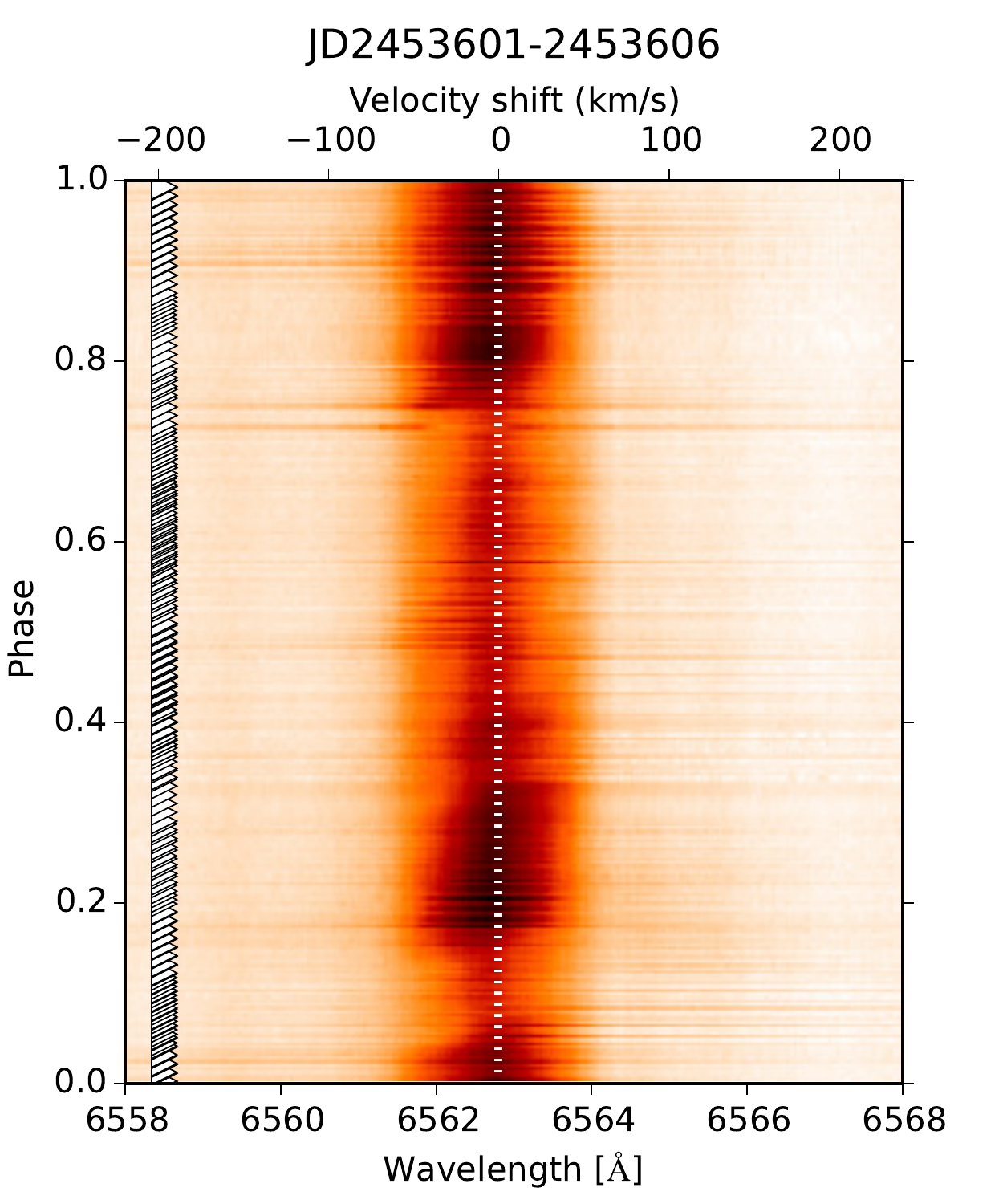}  
	\includegraphics[width= 0.32\textwidth]{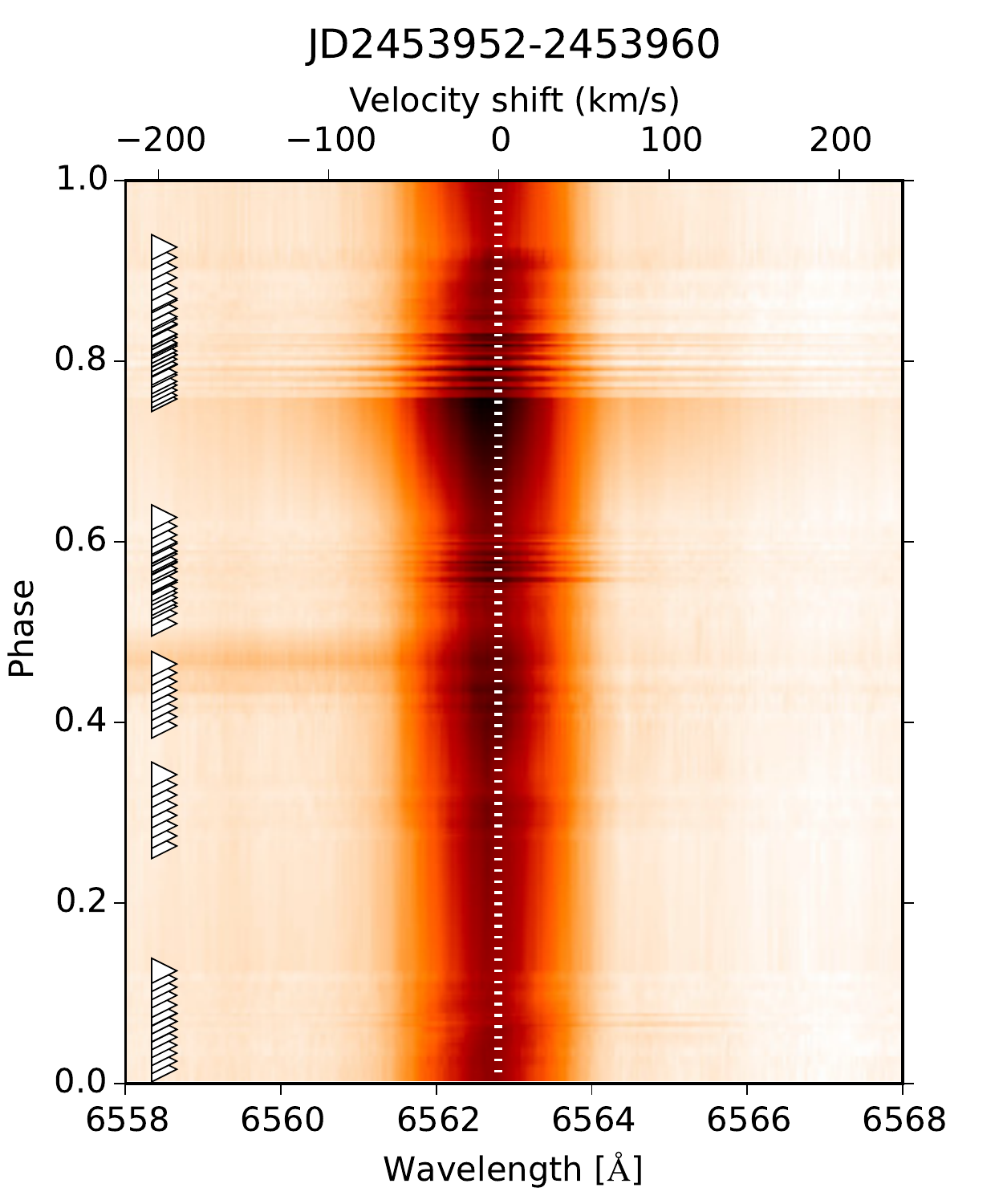}  
	\includegraphics[width= 0.32\textwidth]{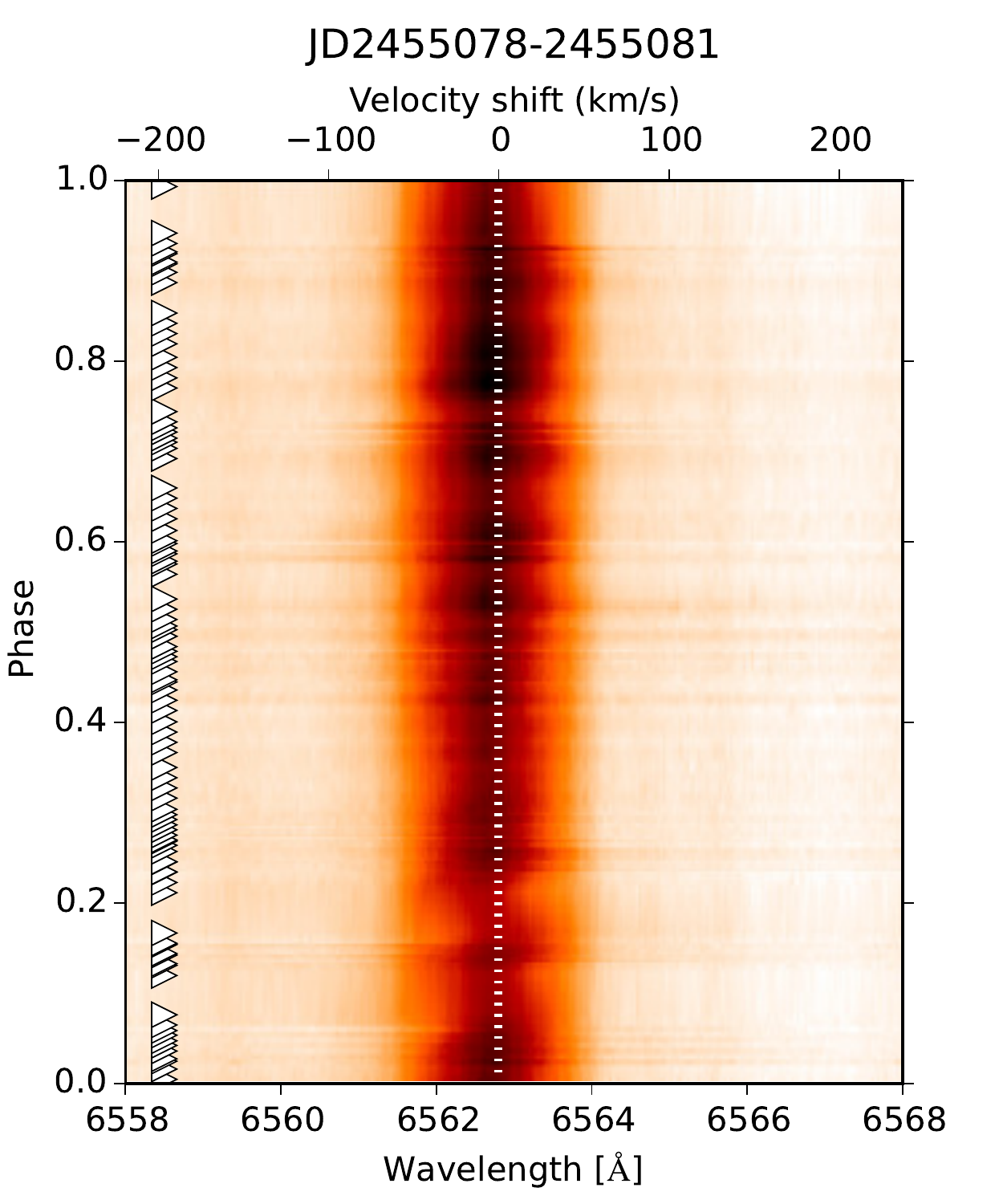}
	\includegraphics[width= 0.32\textwidth]{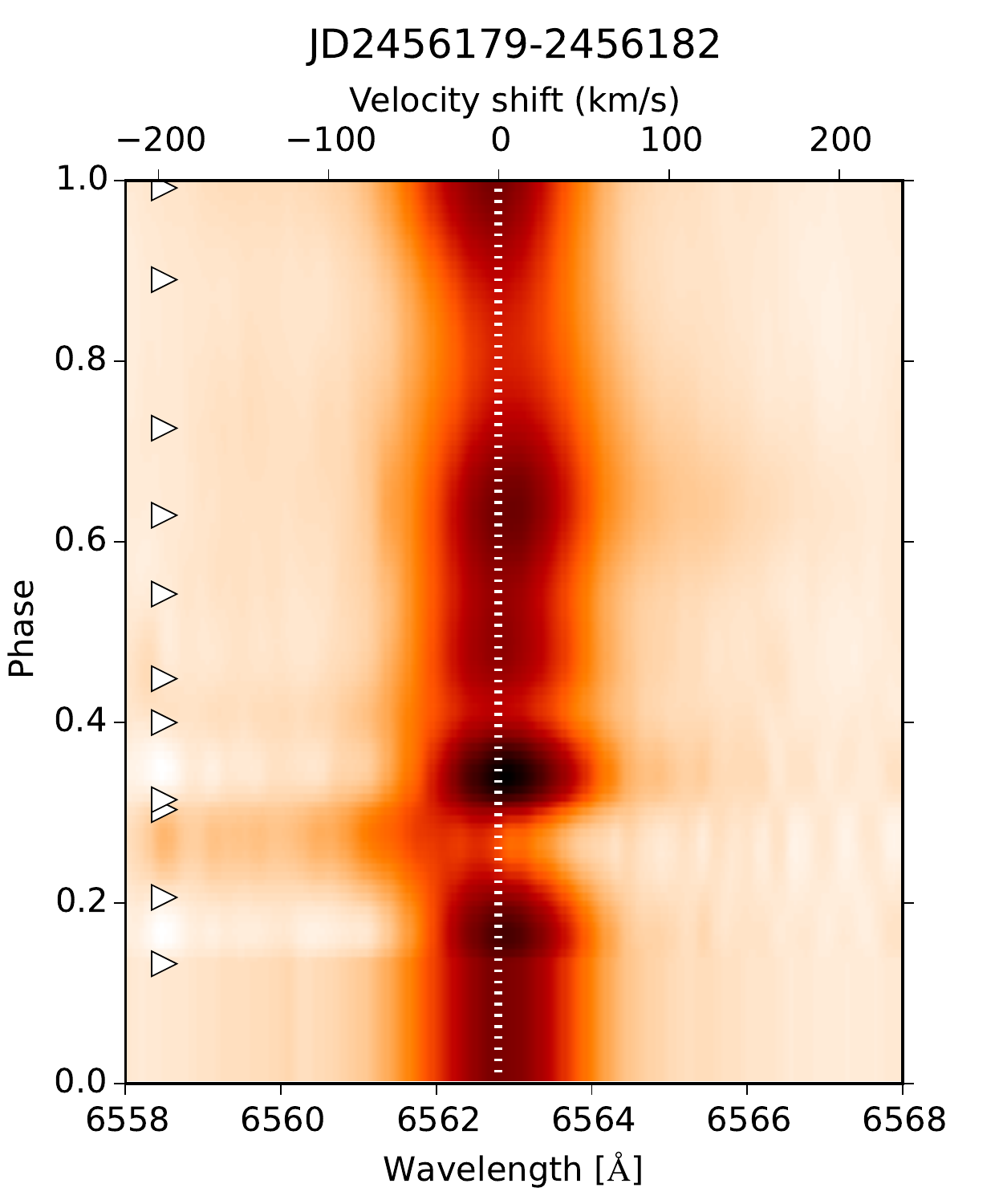}
	\includegraphics[width= 0.32\textwidth]{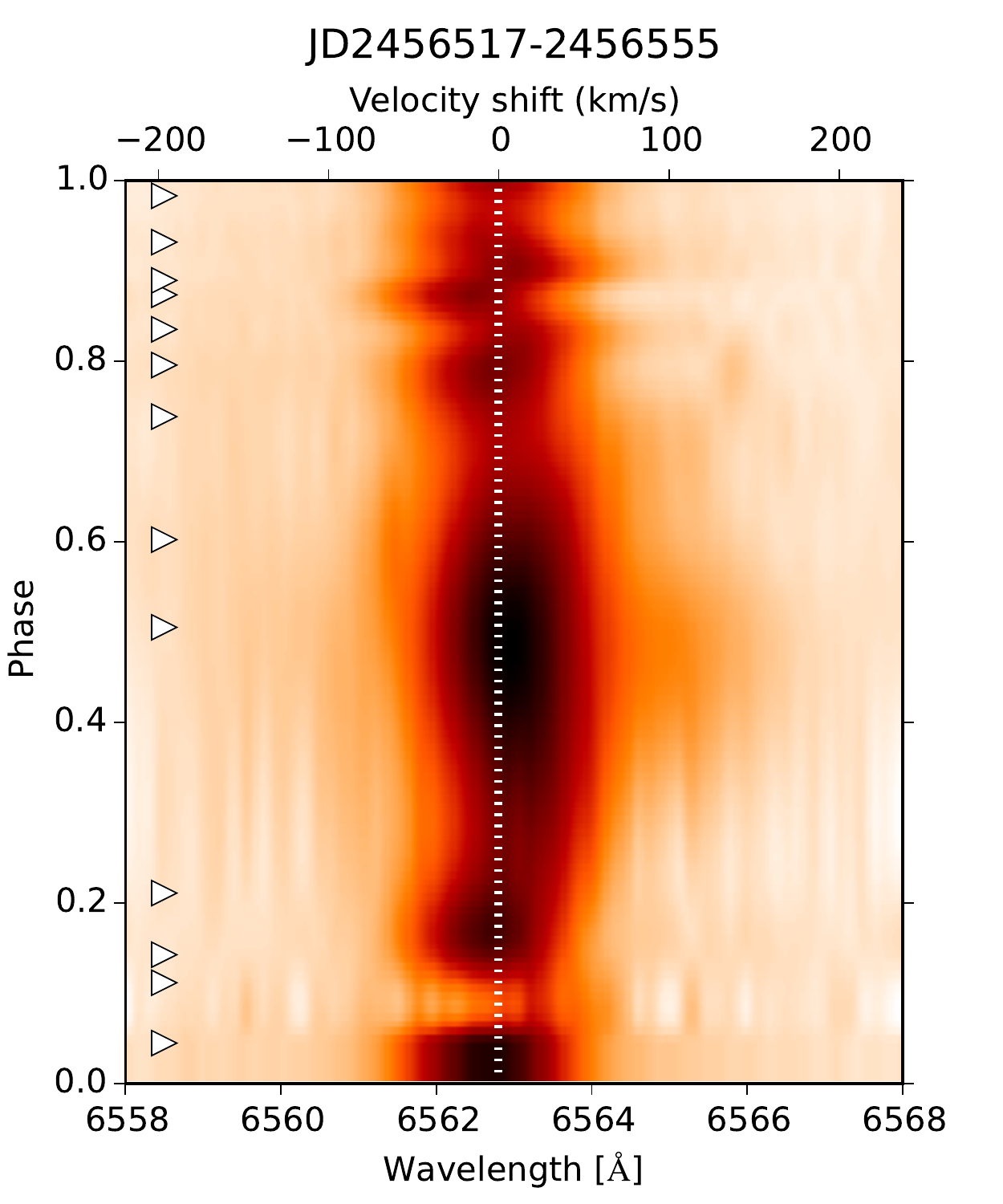}
	\caption{Phased dynamic \HA{} spectra from 2005, 2006, 2009 (CFHT observations, top row), 2012 and 2013 (RCC observations, bottom row). Triangles mark the rotational phases of the individual spectra, dotted line shows the 0 \,km\,s$^{-1}$ velocity shift.}
	\label{fig:dynspecph}
\end{figure*}

\begin{figure*}
	\centering
	\includegraphics[width= 0.32\textwidth]{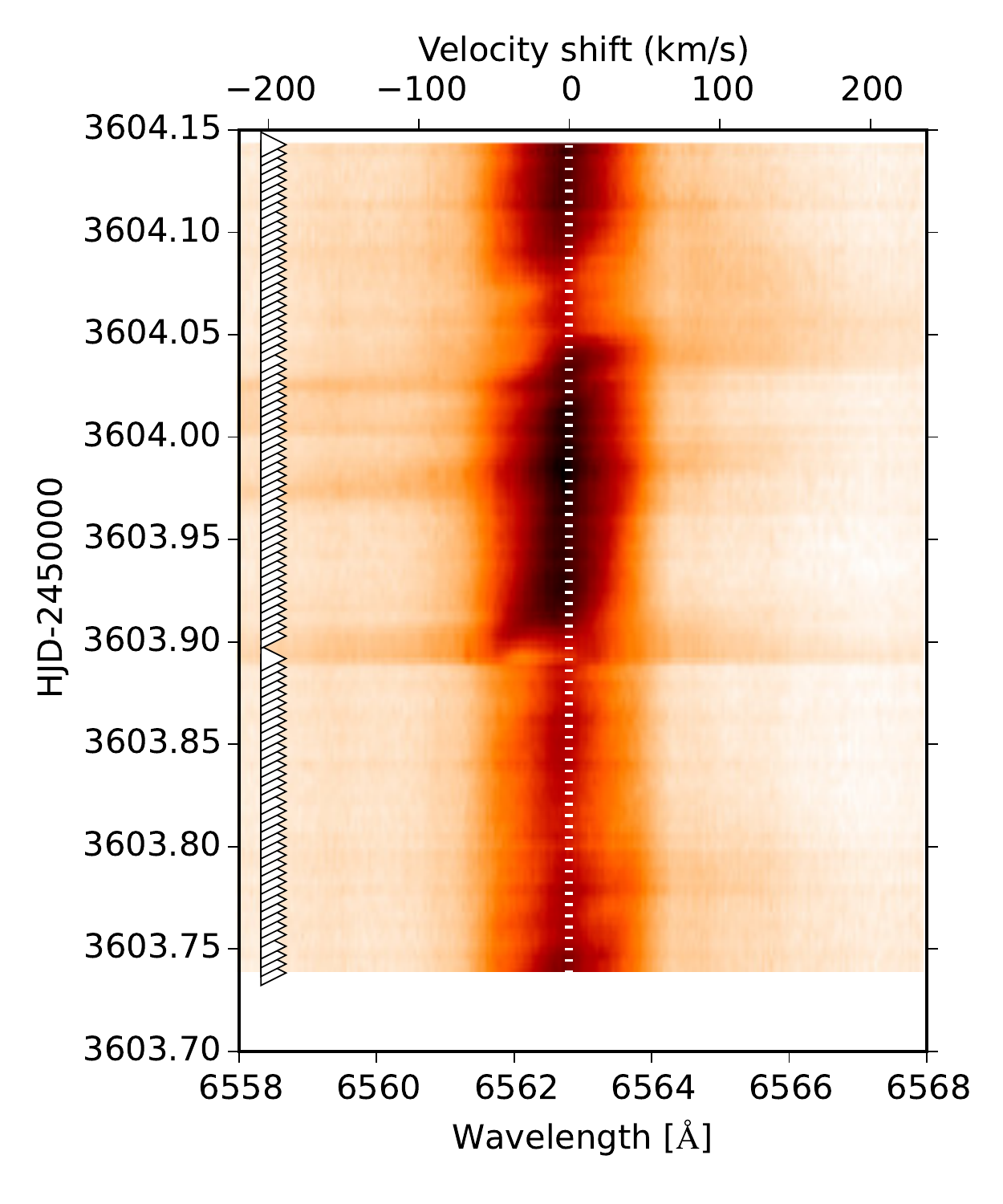} 
	\includegraphics[width= 0.32\textwidth]{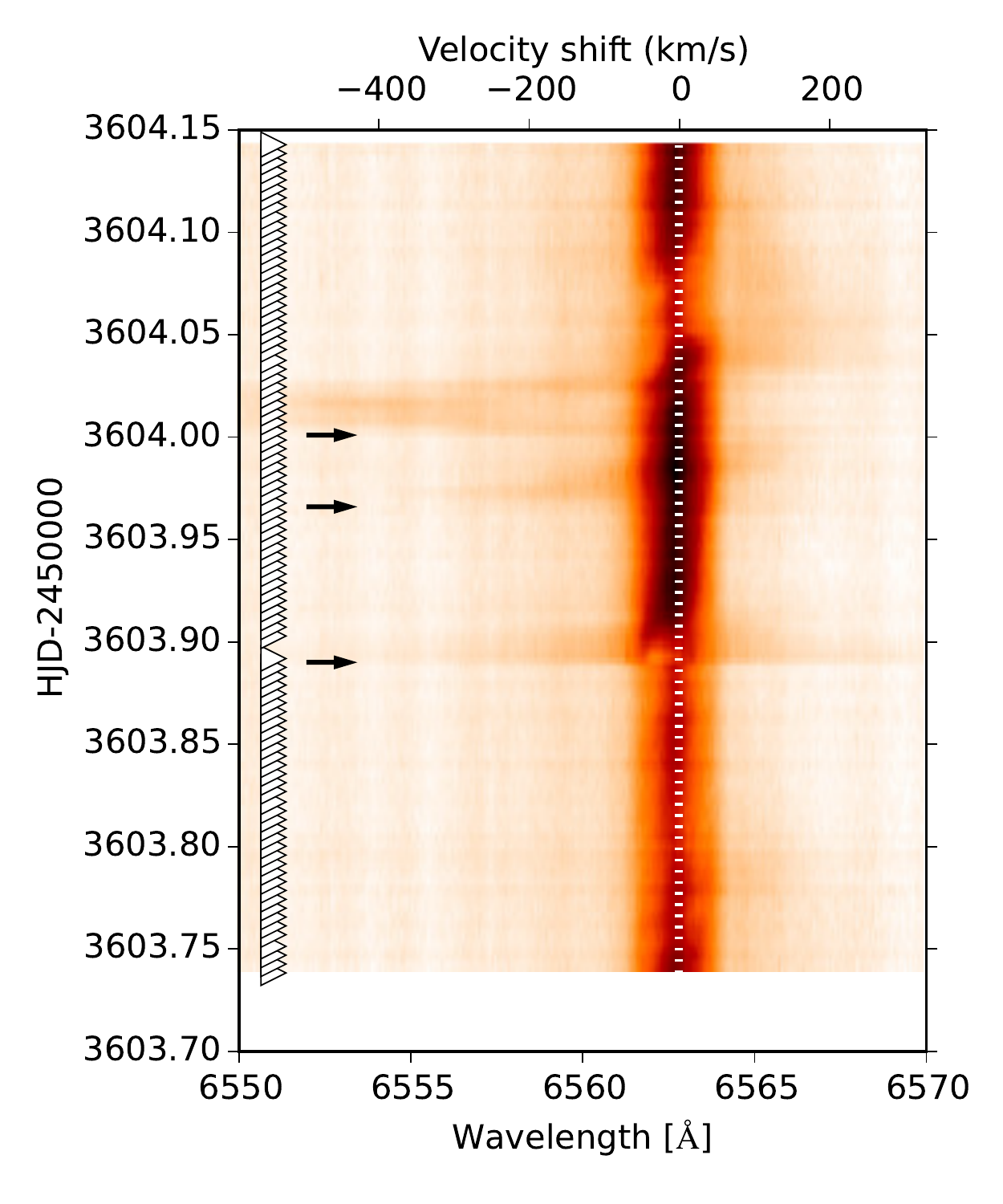}  
	\includegraphics[width= 0.32\textwidth]{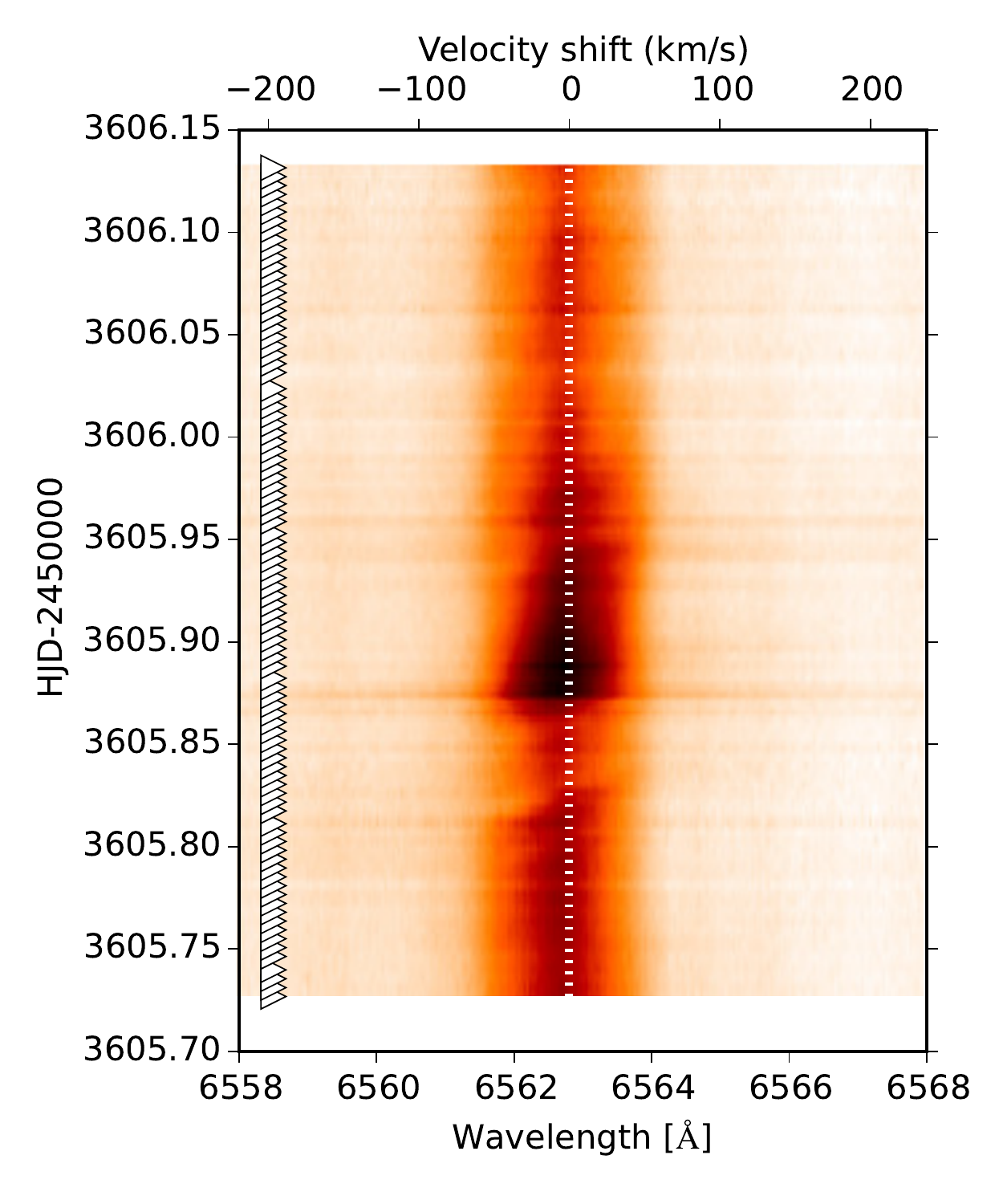}  
	\caption{Dynamic \HA{} spectra of \vpeg{} from 2005, showing strong flare events. The first two plots show the same data with different wavelength scales. This way the CME signature --  the blue wing enhancement -- can be also seen (their start indicated with arrows in the second plot).}
	\label{fig:dynspecjd}
\end{figure*}

\subsection{Flare energies}
\label{sect:flareenergy}
\begin{figure}
	\centering
	\includegraphics[width=0.5\textwidth]{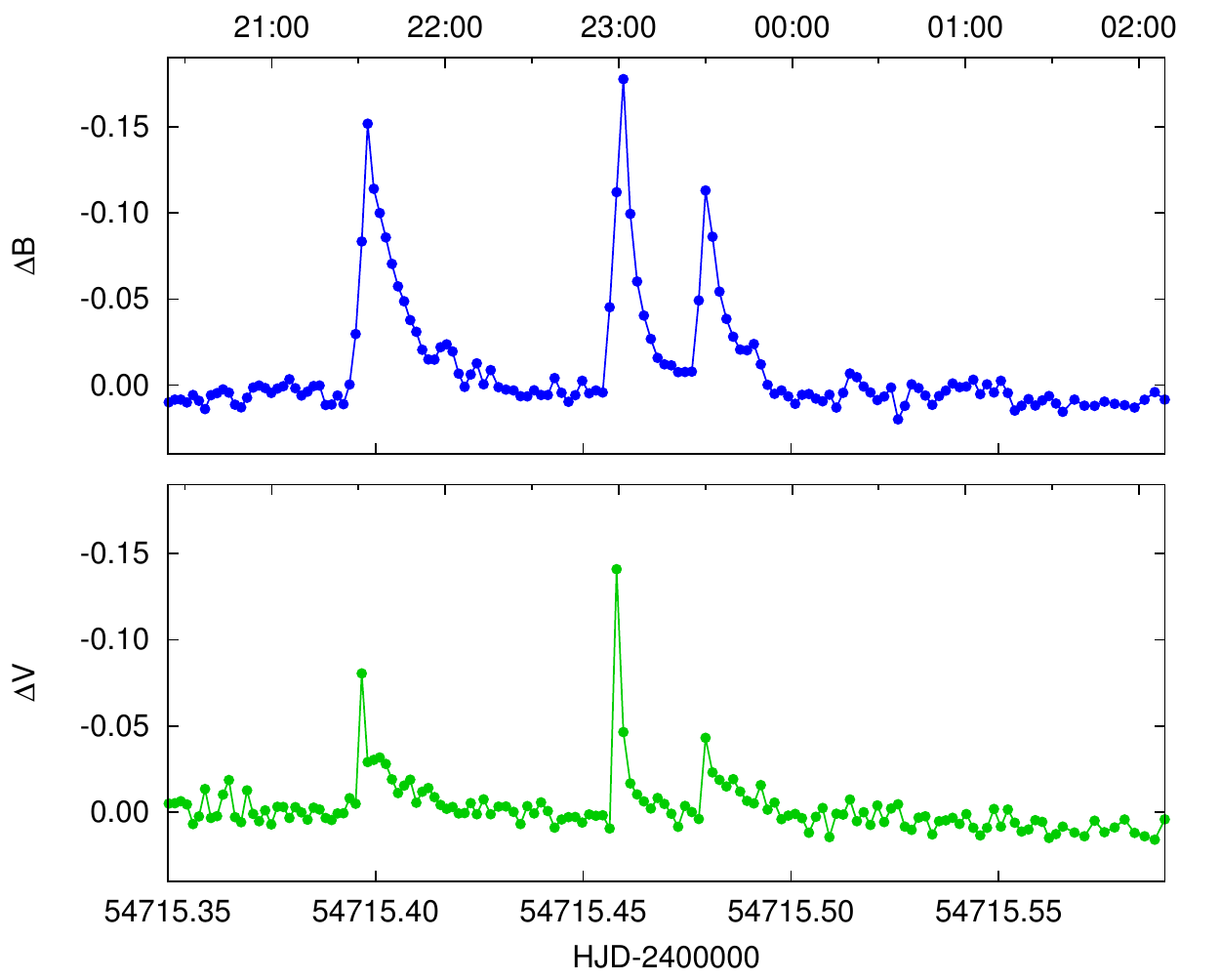}
	\caption{$\Delta B$ and $\Delta V$ light curve of the complex flare event from HJD 2454715.}
	\label{fig:bigflare}
\end{figure}

We selected a few well-observed flare events which were detected at least in two colours. There were deviating points caused by flares also in $R_C$ and $I_C$ passbands, but these detections were too weak for modelling. 
We estimate flare energies in different colours following the method in  \cite{flareenergy}. Derived values are listed in Table \ref{tab:flareenergy}.
Flare energy estimations are carried out as follows. First, to derive the relative flare energy for a given event, the $\frac{I_{0+f}}{I_0}-1$ expression
is integrated over the flare duration, where $\frac{I_{0+f}}{I_0}$ is the ratio of the flaring and the quiescent intensities at the selected photometric band. This quantity, defined by Eq. 2 in \cite{1972Ap&SS..19...75G}, is often mentioned in the literature as equivalent duration. The total integrated
flare energy is then obtained by multiplying the relative flare energy by the basal (i.e., quiescent) stellar flux (see \citealt{flareenergy} for a more detailed description).
Flare energy values are estimated only for those events, which are covered sufficiently by multicolour observations.

Studies dedicated to photometric monitoring of dMe stars such as EV~Lac and AD~Leo \citep{1997A&A...327.1114L,2012NewA...17..399D} detected several flares in $B$-band with energies up to $10^{33}$\,erg, which fits very well to the determined flare energies of \vpeg . Moreover the most energetic solar flares show total energies up to $10^{33}$\,erg \citep{2012JGRA..117.8103S,2010NatPh...6..690K}. As one can see from Table~\ref{tab:flareenergy}, there are flare energies determined from $B$-band photometry, which show similar values, underlining the violent and energetic nature of dMe star flares.

The energy ratio emitted in the different colors are about the same for the first event of a triple outburst at HJD 2454715.394, and for the single events at HJDs 2455059.362 and 2455087.276, being 1.37, 1.41, 1.41 for  $E_B/E_V$, respectively. However, in case of the triple event at 2454715 (see  Fig. \ref{fig:bigflare} for a detailed light curve) the energy ratio $E_B/E_V$ is decreasing during the three consecutive outbursts (separated by about 1.4 and 0.5 hours) as 1.37, 1.28 and 1.10. The emitted energy decreases slower in $V$ than in $B$ passband.  This result strongly suggests that the three consecutive eruptions are related.  Such events are called sympathetic flares which were observed many times on the Sun, where the origin of the flares on the solar surface is well seen. One such interesting example is the 2010 August 1 event on the Sun where three prominence eruptions were observed in the same active region consecutively, with about 6 and 12 hours difference. The details of this event and its theoretical modeling is found in \cite{2011ApJ...739L..63T}, see also \cite{2007ApJ...656L.101A} and cf. \cite{flareenergy}. Briefly, these flare events originate form successive flux rope eruptions triggered by nearby eruptions, in appropriate magnetic field structure of the solar corona.

The structure of the magnetic fields of M-dwarf stars, on a much smaller surface are, to our best knowledge, simpler than that of the Sun. Concerning stellar sympathetic flares, similar repetitive events to the triple flare of V374~Peg were observed on UV~Cet by  \cite{1995MNRAS.277..423P}. The authors explained the phenomenon by supposing strong dipolar magnetic field on the star, where propagation of MHD waves between the poles through the corona and chromosphere triggers the eruptions at the opposite poles. Such scenario is not unlikely on V374~Peg either. According to Zeeman--Doppler imaging results of \cite{Donati:2006:zdi} and \cite{Morin:2008:zdi} V374~Peg do have a strong dipolar magnetic field, which, similarly to UV~Cet, may help in triggering sympathetic eruptions.

\begin{table}
\caption{Measured flare parameters from different passbands}
\label{tab:flareenergy}
\centering
\begin{tabular}{ccccc}
\hline
\multirow{3}{*}{HJD}& \multicolumn{2}{c}{Equivalent duration}& \multicolumn{2}{c}{Total integrated energy}\\
&\multicolumn{2}{c}{[s]}&\multicolumn{2}{c}{[$10^{32}$ erg]}\\
&$B$ &$V$ &$B$&$V$\\
\hline
\hline
2454715.394 & 114.17 & 39.78 & 4.05& 2.96 \\
2454715.453 & 84.28  & 31.50 & 2.99 & 2.34 \\
2454715.474 & 56.27  & 24.41 & 2.00 & 1.81 \\
2455059.362 & 290.46 & 98.05 & 10.30 & 7.28 \\
2455087.276 & 131.00 & 44.22 & 4.65& 3.29 \\
\hline
\end{tabular}
\end{table}

\subsection{Variations of the \HA{} region}

We obtained spectroscopic observations covering eight years from 2005 August to 2013 August. The last three spectroscopic observing runs -- one with the CFHT, the last two with the 1-m RCC telescope -- have photometric support from the same observing season (cf. Fig. \ref{fig:lcall}). As an example, the \HA{} regions from the RCC observations are plotted in the online Fig. \ref{fig:halpha}.

To study the behaviour of the \HA{} line, we plotted its region as dynamic spectra in Fig. \ref{fig:dynspecph}. In these plots the horizontal and vertical axes represent the wavelength/velocity shift and the phase, respectively. The colour coding corresponds to the intensity of the \HA{} region, i.e., colours darken with line strength. Triangles at the left side of the plots show the phases of the observations. The plot regions at missing phases were interpolated to the closest measurements using spline interpolation. Colour coding is set individually for each plot so the typical variations can be well seen.

\begin{figure}
	\centering
	\includegraphics[width=0.5\textwidth]{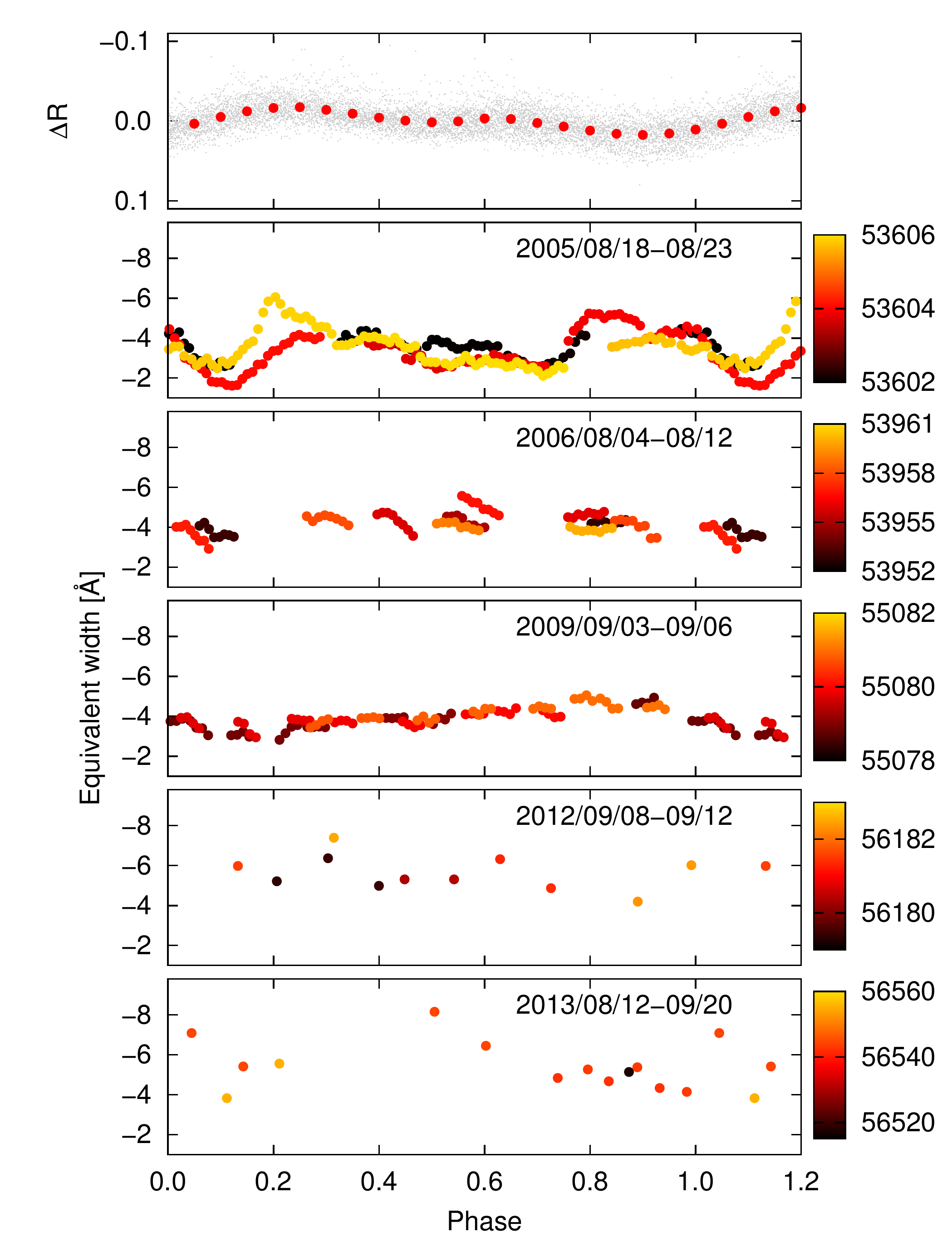}
	\caption{Top plot shows folded, and averaged $\Delta R$ light curve of \vpeg{} including also the observations from 1998 and 2000, after removing long-term trends. The bottom plots show the change of \HA{} equivalent width with rotational phase, colours brightening with time. }
	\label{fig:ew}
\end{figure}

Equivalent width of the \HA{} line was measured for each spectrum using the \verb+splot+  task of IRAF. The variations of the equivalent width with phase was plotted along with a phased light curve (after pre-whitening with long-term trends) for comparison in Fig. \ref{fig:ew} (cf. Fig. 10 of \citealt{Morin:2008:zdi} for a similar analysis of the 2005 observations).  

\begin{figure}
	\centering
	\includegraphics[width=0.5\textwidth]{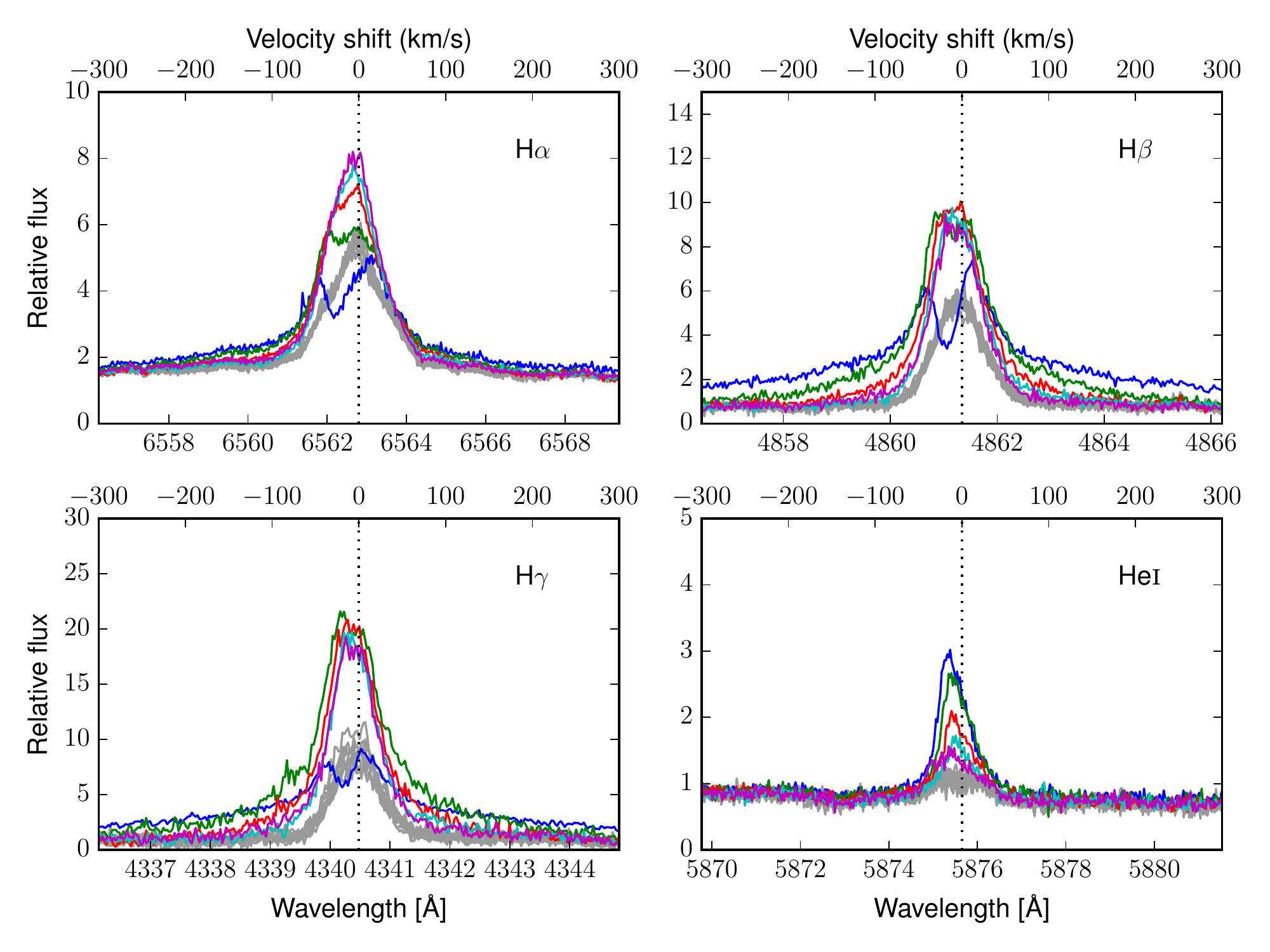}
	\caption{Individual \HA{}, H$\beta$, H$\gamma$, and He {\sc i} spectra from the beginning of the 2453603 flare event. Gray lines show pre-flare spectra, coloured lines show the brightening part of the flare. The spectra were corrected for the radial velocity of \vpeg{} with the value of $V_\gamma=-3$\,\kms{} \citep{Montes:2001:castorgroup}.}
	\label{fig:cme}
\end{figure}

The \HA{} line of \vpeg{} is constantly in strong emission. Obvious rotational modulation can be seen only in the 2009 CFHT spectra, in other cases the phased data are mainly influenced by the flares (this can be also a result of the low resolution). The most powerful \HA{} flares
were observed in 2005. Here, the increase of the \HA{} emission reappears after the consecutive rotations both around phases 0.2 and 0.9 with a small phase shift (see also Fig. \ref{fig:ew}). A possible explanation for this could be long-lived loop systems in the chromosphere that generate flares/coronal mass ejections (CMEs).  

\begin{figure*}
	\centering
	\includegraphics[width= 0.32\textwidth]{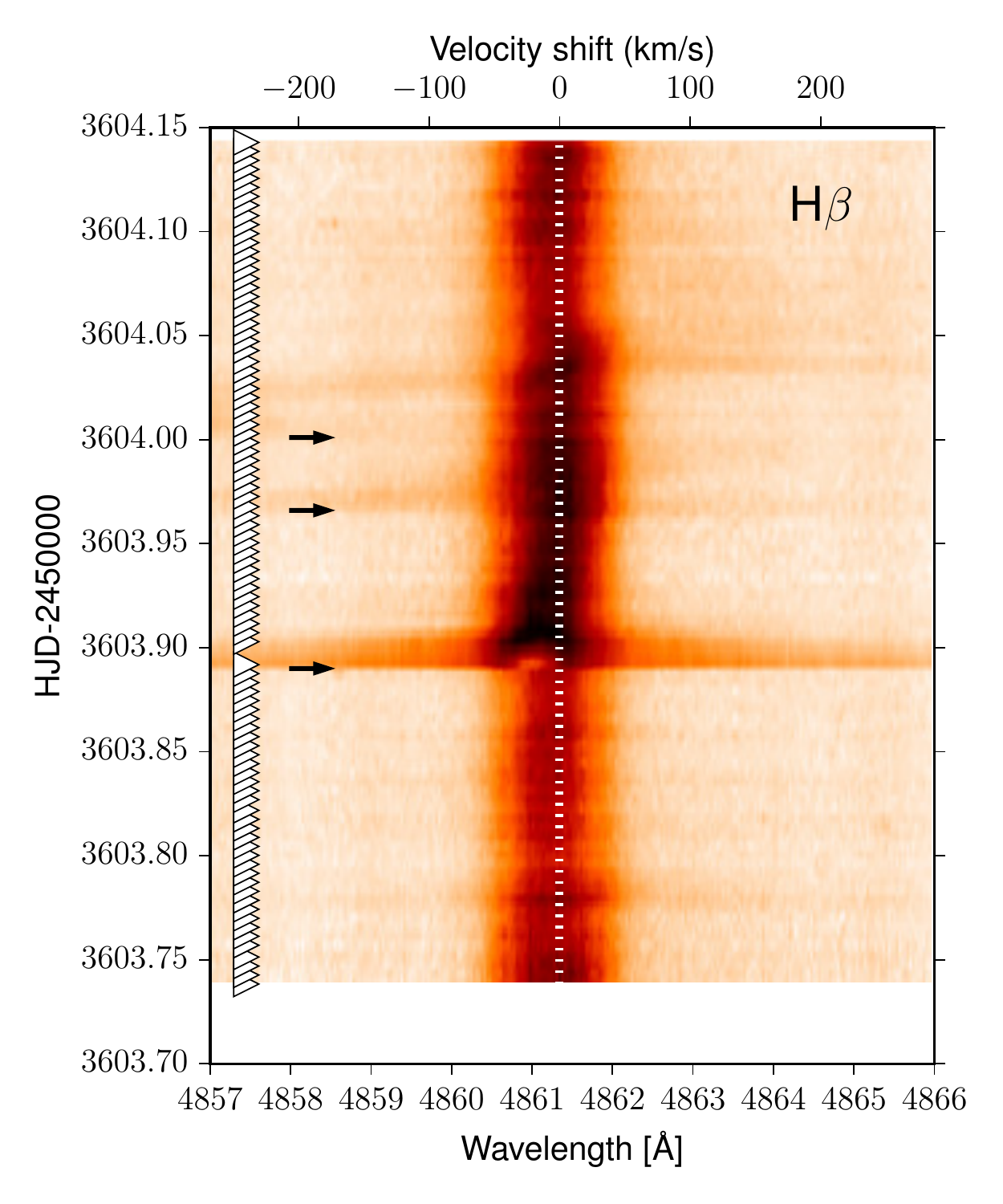} 
	\includegraphics[width= 0.32\textwidth]{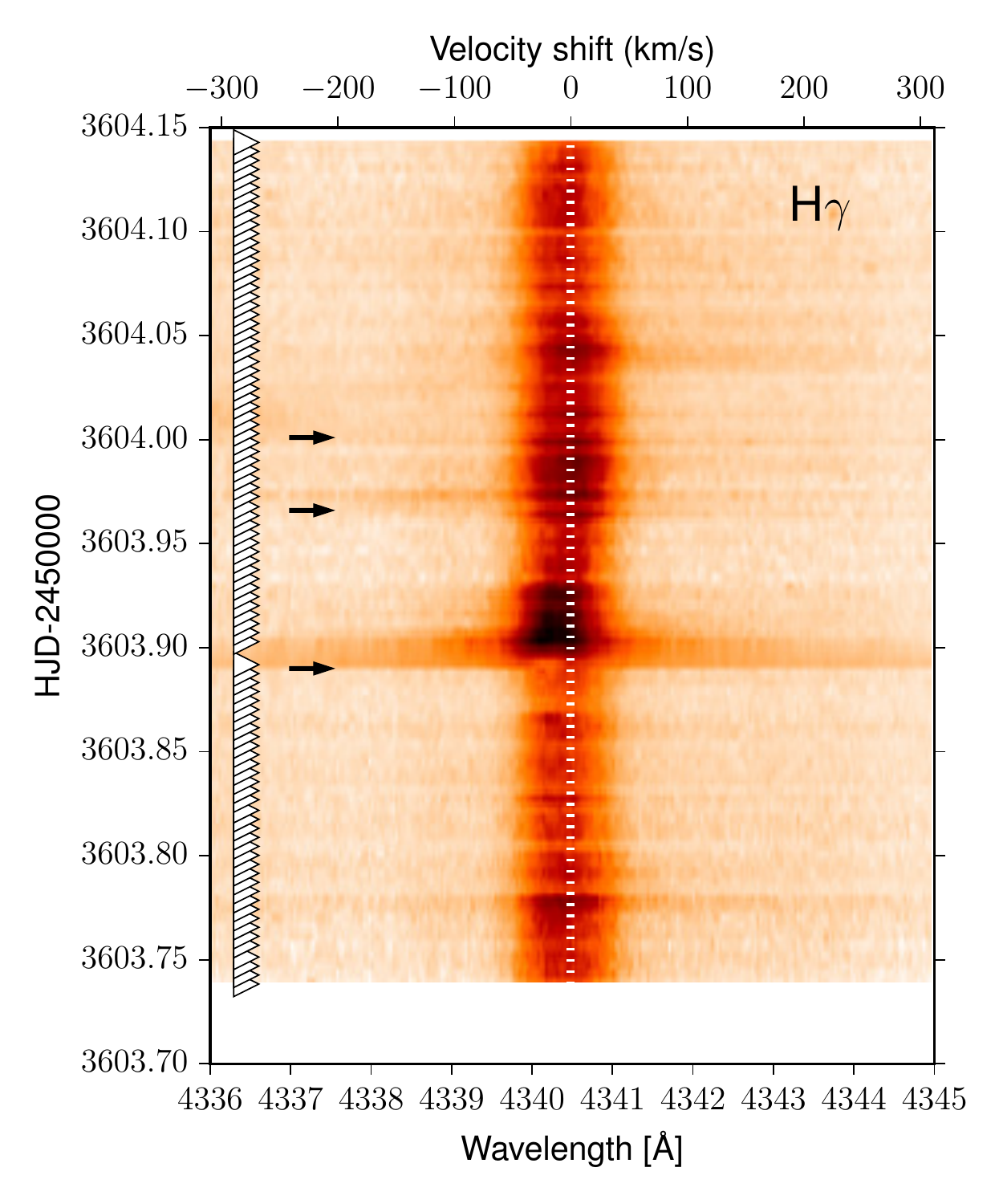}  
	\includegraphics[width= 0.32\textwidth]{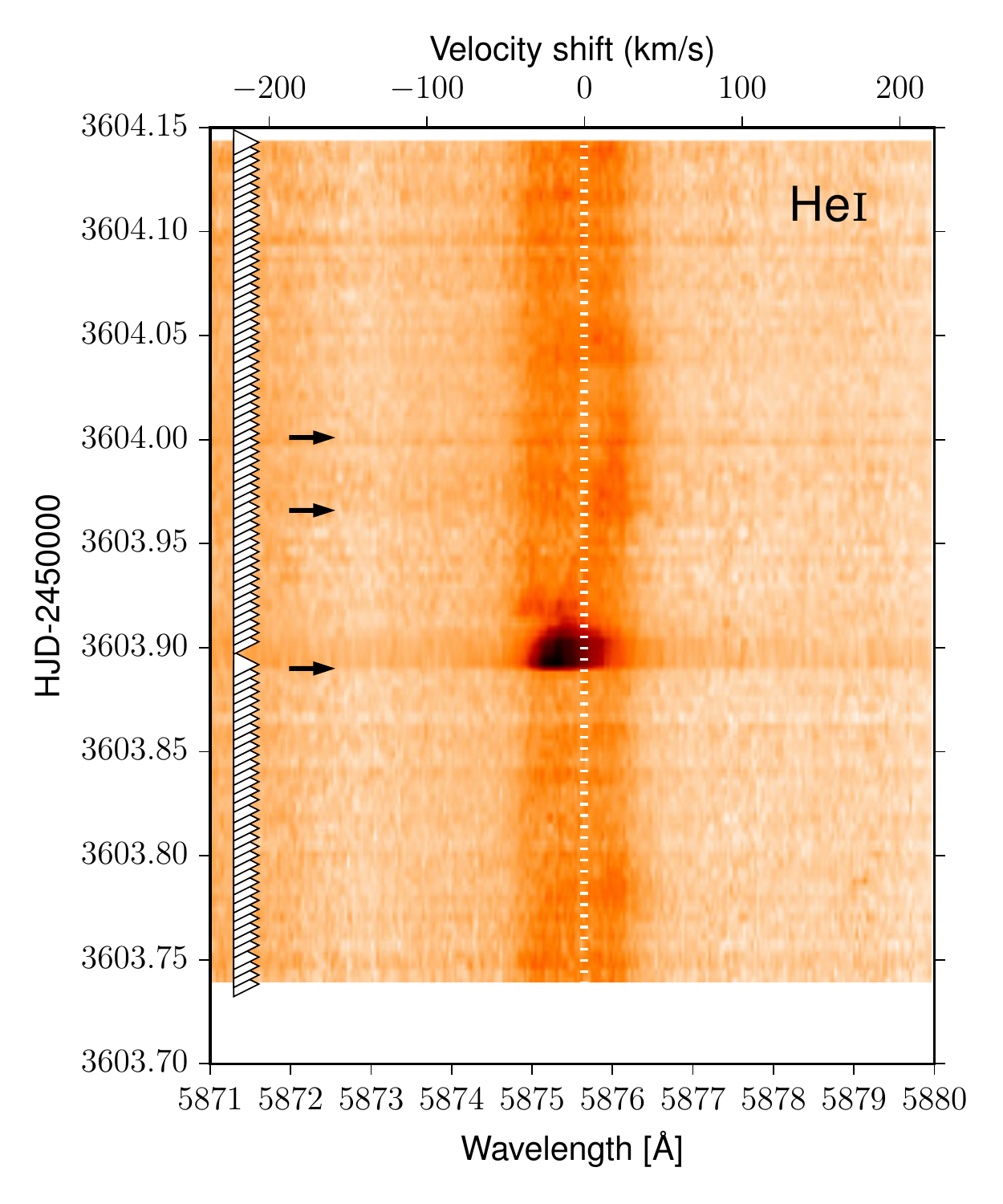}  
	\caption{Dynamic H$\beta$, H$\gamma$, and He {\sc i} spectra of \vpeg{} from HJD 2453603, showing strong flare events. Arrows mark the start of the blue wing enhancements(see text).}
	\label{fig:dynspecph-otherlines}
\end{figure*}
\begin{figure}
	\centering
	\includegraphics[width=0.45\textwidth]{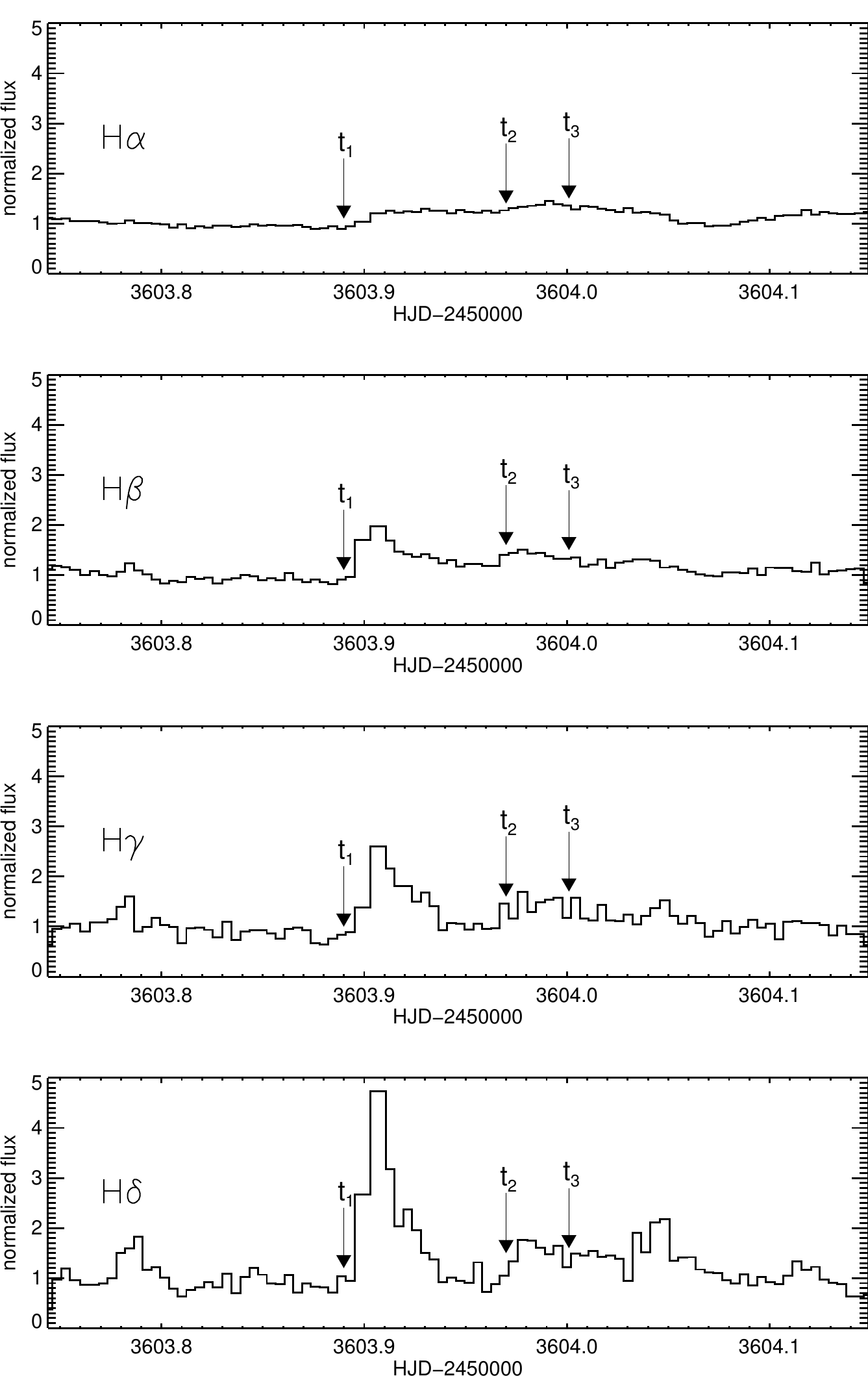}
	\caption{Light curves for the HJD 2453603 event for H$\alpha$, H$\beta$, H$\gamma$, and H$\delta$. The flare, which can be seen easily, is pronounced in H$\beta$, H$\gamma$, and H$\delta$, and much weaker in H$\alpha$. The starting times of the BWE are marked with arrows.}
	\label{fig:ha-lc}
\end{figure}

On HJD 2453603 (see dynamic spectra in Fig. \ref{fig:dynspecjd}, H, He lines in Fig. \ref{fig:cme} and intensity curves for the Balmer lines in Fig. \ref{fig:ha-lc}), one can see 3 distinct blue wing enhancements with different durations and projected velocities, spanning over more than three hours. The detailed chronology of these events is described in Appendix \ref{sect:cmedesc}.

Interestingly, the overall level of the \HA{} line is quite low at the time of the first, double (or triple) peaked spectrum (see the spectrum plotted in blue in Fig. \ref{fig:cme}). This could be similar to the pre-flare "dips" observed by \cite{2014MNRAS.443..898L} in the young open cluster Blanco-1, although this feature -- if real -- is only seen in one spectrum, and only in the \HA{} region. Unfortunately, these events are very short (from a few seconds to a few minutes, see \citealt{2014MNRAS.443..898L} and references therein) and rather weak, thus the 5-minute sampling in our spectra is not optimal for detecting these events.

To measure the flux of the enhancement we select wavelength windows which vary from 6547.0--6561.1\,\AA{} (722--78\,\kms) to 6555.0--6561.1\,\AA{} (357--78\,\kms) from which we subtract the continuum flux measured in the same windows. The continuum flux level was adopted from the M4V star GJ699 taken from \citet{2004A&A...414..699C}. As the blue-wing enhancement occurred during a flare on V374~Peg it is difficult to disentangle the contribution from the flare and the CME. We expect that the contamination of the blue wing of H$\alpha$ from the flare occurs at small velocities close to H$\alpha$ line center. With the usage of fixed windows we underestimate the total flux caused by the ejection. The estimation of the CME mass depends on the measured flux. However, with the method we use for the estimation of the CME mass (see next paragraph) we are able to determine a lower limit to the CME mass only.

From the \HA{} blue wing flux enhancement one can estimate the mass of the CME. We use
\begin{equation}
M_\mathrm{CME} \ge \frac{4 \pi d^{2} F_{em} (N_{tot}/N_{i}) m_{H} \eta_{OD}}{h\nu_{j-i} A_{j-i}}
\label{mcme}
\end{equation}
adopted from \citet{1990A&A...238..249H}. Eq.~\ref{mcme} is comprised of the distance of the star $d$, the flux of the emission feature $F_{em}$, the total number of hydrogen atoms $N_{tot}$, the number of hydrogen atoms in an excited state $i$, $N_{i}$, the opacity damping factor $\eta_{OD}$, the mass of a hydrogen atom $m_{H}$, the Einstein coefficient for transition $A_{j-i}$, frequency $\nu$, and Planck's constant $h$. As we have no estimation for the ratio $N_{tot}/N_{3}$ (corresponding to \HA{}) we follow the method from \citet{2014MNRAS.443..898L}. First they used the Balmer decrement ($BD$) to scale the measured H$\alpha$ flux to H$\gamma$, i.e. $F_{em,\gamma}=F_{em,\alpha}/BD$, with $BD=3$ valid for solar and stellar flares (Butler et al. 1988). Then the parameters of the H$\gamma$ line can be used in Eq.~\ref{mcme}, including the value $N_{tot}/N_{5}\sim2\times 10^{9}$ given by \citet{1990A&A...238..249H} which is valid for temperatures of about 20\,000~K and densities of $10^{10}-10^{12}~\mathrm{cm^{-3}}$ and was derived from NLTE radiative transfer modelling. Furthermore, we adopt $\eta_{OD}=2$ to account for optical thickness of the lower Balmer lines \citep{1990A&A...238..249H}.

According to the above described estimation of the CME mass, we find a minimum CME mass in the order of 10$^{16}$g. This value is comparable to the most massive solar CME masses  reaching 10$^{17}$g (see e.g. \citealt{2010ApJ...722.1522V}).

To estimate the expected number of CMEs above $10^{16}$ g mass, we make use of the method applied in \citet{2014MNRAS.443..898L}. This method uses flare rates predicted by the scaling of flare rate with stellar X-ray luminosity from \citet{2000ApJ...541..396A} together with a CME mass-flare energy relation from the Sun \citep{2013ApJ...764..170D} to estimate stellar CME rates from stellar X-ray luminosities. Adopting $\log L_X\approx28.4$ erg\,s$^{-1}$ for V374 Peg \citep{1999A&AS..135..319H} we infer 15--60 CMEs per day with a mass $>10^{16}$ g for a range of flare power law indices $\alpha$ of 1.8--2.3 \citep[][their Fig. 5]{2014MNRAS.443..898L}. These numbers are larger than what has been observed in this study (1 event in 10 h), but the detectability of stellar CMEs is affected by projection effects (random directions of ejection, not always leading to sufficiently blue-shifted signals). Moreover, the extrapolation of flare--CME relations known from the Sun to young active stars is still debated \citep{2013ApJ...764..170D}. 
In section \ref{sect:CMEdiscussion} we discuss these results with respect to the so far detected stellar CMEs from literature, which are sparse.

 In the case of the HJD 2453605 flare event, no significant velocity shifts can be seen that could indicate CMEs.

\section{Discussion}   

\subsection{Connection between the photosphere and chromosphere}
The connection between the chromospheric and photospheric activity on the Sun is known for a long time. Such connection was also observed on other active stars, e.g., in the case of the dM1--2 type EY Dra \citep{2010AN....331..772K}, the fully convective K7-type T Tauri star TWA 6 \citep{2008MNRAS.385..708S} and other BY Dra type stars \citep{2005AJ....130.1231P}. 
To study the behaviour of the chromosphere, the \HA{} region is the most widely used proxy for M dwarfs. 
In the case of a Sun-like connection between the photosphere and chromosphere we would expect anticorrelation between the light curve and the \HA{} equivalent width, i.e., the chromospheric activity would increase at those phases, where we see the dark spots in the light curve. The case is not necessarily that obvious, since bright photospheric structures on the Sun, as well as prominence-like features are seen better off-limb rather than when projected against the stellar disc \citep{Hall:1992bx}. 
In Fig. \ref{fig:ew} the \HA{} equivalent widths from different epochs are plotted along with 
the phased $\Delta R$ light curve. This dataset has the longest time-span of 16 years, and shows stable light curve all the time, before and after the spectroscopic observations were taken.
Unfortunately the spectroscopic results from the RCC telescope (in 2012 and 2013) are quite inconclusive, as the size of the telescope did not allow to use short exposures, thus the shorter-term variations caused by possible flares cannot be temporally resolved. In the case of the CFHT data (between 2005 and 2009) an obvious anticorrelation can be seen  in 2009 September. Without any indication of flares, just a rotational modulation was apparent. Interestingly, this is the part of the photometric observation, where frequent strong flare activity was observed, there are photometric observations three weeks before, and one week after these spectra were obtained. In the case of the 2005 and 2006 spectra, no global trends are seen, just variations caused by flares, especially in 2005. Here, the increase of the \HA{} activity seems to be concentrated around phases 0.2 and 0.8--1.0, the latter being the part where the light curve shows the faintest state. 

\begin{figure}
	\centering
	\includegraphics[width=0.48\textwidth]{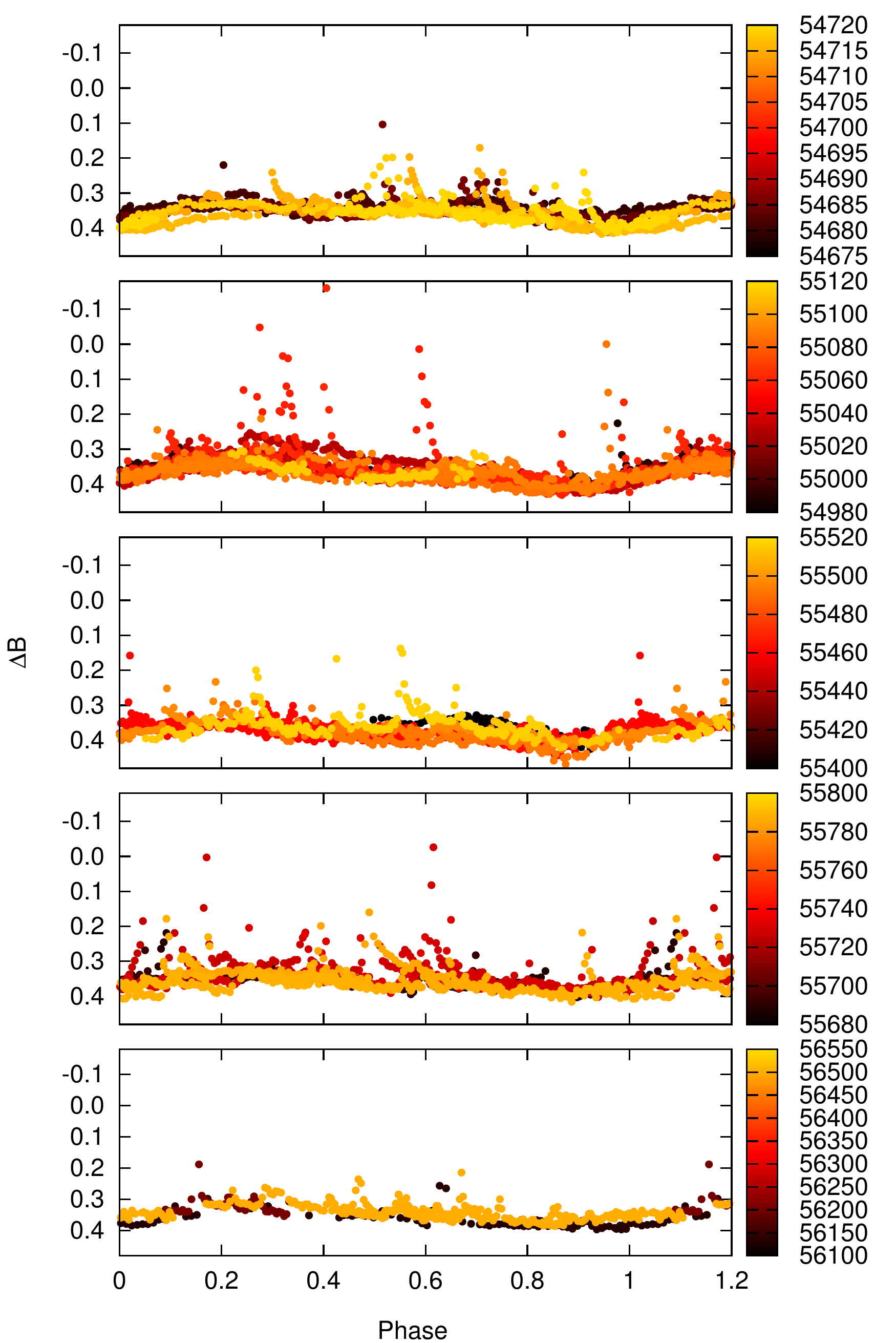}
	\caption{Phased $\Delta B$ light curves as used for flare statistics. Colours brighten with HJD. 
	Note that the second and fourth plot do not show the strongest flares.}
	\label{fig:flarelc}
\end{figure}
In Fig. \ref{fig:flarelc} folded $\Delta B$ light curves are plotted in those epochs that were used for flare statistics (see Sect. \ref{sect:flarestat}). The global shape of the light curve, i.e., the large-scale spot configuration does not change in this time significantly. Flares can be seen in all cases at every phase, however the strongest flares seem to be more concentrated around phase 0.3--0.7 (see also Fig. \ref{fig:flarehist}).

\subsection{Photometric variability}
We found that the light curve, thus the overall surface structures are stable for at least 16 years on \vpeg{} (cf. Fig. \ref{fig:ew}, top panel). This agrees with the finding of \cite{Morin:2008:zdi} who found that the magnetic configuration did not change on a yearly time scale. The star furthermore seems to rotate as a rigid body, or has a very weak differential rotation as we found from the Fourier analysis (see Sect. \ref{sect:fourier}), in agreement with the results of \cite{Donati:2006:zdi} based on Zeeman--Doppler images. According to the theoretical model of \cite{2011AN....332..933K}, we would expect $\alpha \approx 0.008$ for a star of $0.3M_\odot$ and $P_\mathrm{rot}=0.5\,d$.  Our findings are in good agreement with recent theoretical dynamo model for fully convective stars of \cite{2015ApJ...813L..31Y}  which suggests that these stars have stable magnetic fields, very small differential rotation, polar spots, and active regions distributed throughout the stellar phase (implying flare distribution independent of rotation phase). 

However, besides this stable spot configuration, the light curve is not constant at all: small changes can be seen continuously on a nightly timescale (see Fig. \ref{fig:flarelc}). These variations are mainly seen between phases 0.3--0.6. These minute intensity variations could be caused by newly born and decaying small spots in the same active nest, similarly to the emerging flux ropes in active nests on our Sun.

\subsection{CMEs}
\label{sect:CMEdiscussion}
The CME/flare detected on V374~Peg in the CFHT data on HJD 2453603 represents a unique event compared to the ones from literature. First of all the length of the  observations (more than 10 hours of continuous monitoring) enables the comparison of variability before and after the event.  Stellar CMEs are sporadically observed phenomena, and their detection so far presented in literature happened by chance, except the event presented by \citet{1997A&A...321..803G}, who performed a dedicated search for stellar activity through spectrophotometry in \HA . 
In literature only single events, which the authors have assigned to stellar CMEs, can be found. \citet{1990A&A...238..249H} detected  the fastest CME so far on a late-type main-sequence star AD~Leo in the optical, which showed a maximum projected velocity of $\approx$ 5800~km~s$^{-1}$. Further events in the optical domain were presented by \citet{1994A&A...285..489G} on the dMe star AT~Mic, \citet{1997A&A...321..803G} on a dM0 weak-line T-Tauri star, and  \citet{2004A&A...420.1079F} on an old dM star.
Stellar CME events in the UV/FUV were presented by \citet{2001ApJ...560..919B} on the pre-cataclysmic binary V471~Tau, and \citet{2011A&A...536A..62L} again on AD~Leo. Interestingly, all detected CMEs with the method of Doppler-shifted emission in optical spectra have been detected on dMe stars to date.

In the case of V374~Peg, we detected three distinct blue-wing enhancements which happened during a flare (cf. the discussion on sympathetic flares in Sect. \ref{sect:flareenergy}), after a quiet period of at least 3 hours (see Figs \ref{fig:dynspecjd}, \ref{fig:cme} \ref{fig:ha-lc}). The chronology of the three events is that at $t_{0}$, BWE1 arises simultaneously with an H$\alpha$ flare. Nearly 2 hours later, BWE2 begins to rise, and finally 50 minutes later the fastest event, BWE3 starts. We calculated an escape velocity for V374~Peg of $v_{e}\approx$580~km~s$^{-1}$ using 0.30 \msun{} 
and 0.34 \rsun{} (see Sect. \ref{sect:mass}). The projected maximum velocity of BWE3 (675\,\kms) is above the escape velocity, this means that during this event, mass was ejected from the star. The two precursor events (BWE1, BWE2) show maximum velocities below $v_{e}$ (each $\approx -350$\,\kms) and therefore can  not be distinctively assigned to mass ejected from the star. If we assume that all three events belong to the same active nest then BWE1 and BWE2 could also be failed eruptions. On the Sun, events have been reported \citep{2006ApJ...651.1238Z} where prominences failed to erupt, while the remaining filament has reformed, being then ejected, and observed as a CME. The observed red wing enhancements after the CME event could be a result of material falling back on the stellar surface, as seen on the Sun.

If we assume that the event is connected with the photospheric active regions, as seen e.g. on the Sun, but also other stars (see e.g. \citealt{2010AN....331..772K}), we can give an estimation of the real velocity of the CMEs. According to the spot models, both  active nests are located around $70^\circ$ latitude (although this is the most uncertain spot parameter), this means that the CME happened roughly along the line of sight assuming an inclination of $70^\circ$ (see \citealt{Donati:2006:zdi}). 
This suggests that the projected velocity could be close to the true velocity of the ejection.
An uncertainty of $\approx10^\circ$ in the inclination or the spot latitude would cause only a difference of 1--2\% in the maximal speed of the ejecta.

The three events span over a time range of $\approx$3~hours, which corresponds to about 30\% of the stellar rotation period. If, as assumed above, the signatures are connected to the same active nest, then the rotation of the nest has a significant influence on the  projected  velocity of the events.  BWE1 started at phase 0.72, while the centre of the large spotted region is at $\lambda\approx340^\circ$, i.e. 0.94 phase. If this eruption happened in the middle of this nest, we would thus only observe $\approx19\%$ of the actual velocity (neglecting the effect of the latitude), for BWE2 and BWE3 this value is 95\% and 98\%, respectively. The large size of the active region ($\gamma\approx42^\circ$ radius) however causes a rather large uncertainty in these estimated values.

 Although V374~Peg has a rotation period of $<$ 0.5 days, its H$\alpha$ profile is not sufficiently rotationally broadened \citep[$v\sin i=36.5$\,\kms, see][]{2008MNRAS.390..567M} to detect absorption features as signatures of co-rotating clouds \citep{1989MNRAS.236...57C} which are commonly interpreted as stellar analogs of solar prominences.
 
The ejected mass at BWE3 which we deduced to be in the range of $\approx$10$^{16}$g fits to the mass of massive solar CMEs. However, the method to determine the mass of stellar CMEs used in this paper is probably accurate only within an order of magnitude. Non-Local Thermal Equilibrium (NLTE) modelling is necessary to self consistently calculate the radiative transfer from which a more accurate mass may be determined.

The theoretical CME rate which we have deduced based on the method presented in \citet{2014MNRAS.443..898L}, relies on the solar CME mass--flare energy relation \citep{2012ApJ...760....9} and the empirically determined flare power-law from \citet{2000ApJ...541..396A}. Depending on the flare index $\alpha$ the obtained CME rate is 15--60~CMEs (M>M$_{c}$, M$_{c}$=10$^{16}$\,g) per day.
However, only a fraction of this number can be observed due to projection effects when assuming that CMEs can be ejected anywhere from the star.

The stellar parameter on which the theoretical CME rate depends on is the stellar X-ray luminosity as indicator for activity. V374~Peg has a $\log L_x$ of 28.4\,erg\,s$^{-1}$ \citep{1999A&AS..135..319H} which is roughly one order of magnitude higher than for the Sun during activity cycle maximum \citep{2000ApJ...528..537P}. The observed CME rate of the Sun considering a whole cycle \citep{2009ApJ...691.1222R} lies between 0.5 and 8 CMEs per day on average which is lower than the theoretical CME rate of V374~Peg, as expected (although it is possible that the reason for this discrepancy is that we simply cannot observe most of the stellar CMEs). One has to keep in mind that the solar--stellar analogy might work for solar-like stars, the extrapolation of solar knowledge to dM stars has to be done very carefully.
The question if young stars exhibit also more massive CMEs than the Sun is still an open issue. 
 
However, the extrapolation of solar CME parameter distributions is  the only possibility at the moment to gain knowledge on CME frequency on stars others than the Sun, since no observationally determined CME parameter distributions of late-type main-sequence stars exist.

\section{Summary}     
We summarised stellar parameters of \vpeg{} in Table \ref{tab:params}.
\begin{table}
\caption{Parameters of \vpeg .}
\begin{tabular}{lll}
\hline
\hline
Spectral type & dM4 & \cite{1995AJ....110.1838R}  \\
Age & 200\,Myr & \cite{Montes:2001:castorgroup} \\
Distance & 8.93\,pc &\cite{hipparcos} \\
$v \sin i$ & 36.5\,\kms & \cite{Morin:2008:zdi}\\
Assumed inclination & $70^\circ$& \cite{Donati:2006:zdi}\\
Mass& 0.30\msun & $^\dagger$ \\
Radius& 0.34\msun & $^\dagger$ \\
$P_\mathrm{rot}$ &$0.44570$\,d &  $^\dagger$ \\
Effective temperature& 3400\,K& $^\dagger$\\
Spot temperature & 3250\,K& $^\dagger$ \\
\hline
\end{tabular}

\vspace{2mm}
$^\dagger$ this paper
\label{tab:params}
\end{table}

\begin{itemize}
\item The light curve is stable over about 16 years, that confirms the previous indications of a very stable magnetic field;
\item {Besides the stable spot configuration, small changes can be seen on a nightly timescale, possibly caused by newly born and decaying small spots in the same active nest;}
\item According to Fourier analysis, there is only one significant peak in the power spectrum corresponding to the stellar rotation and an additional small amplitude signal close to it;
\item The Fourier spectrum suggests very weak differential rotation (almost solid-body), we found indication of a shear parameter of $\alpha=0.0004$;
\item There is no sign of activity cycles in the light curve;
\item Frequent flaring was observed. The occurrence rate of weak flare seems to be  similar in every season on average, but the frequency of strong flares varies on weekly-monthly timescales within a year;
\item Flares were observed at every rotation phase, the strongest flares seem to be more concentrated around phase 0.3--0.7, i.e. where the light curve indicates a smaller active region 
(or an activity nest on the southern hemisphere that can be only partly seen);
\item In the light curve a complex flare was detected with repeated eruptions. The similar $B/V$ energy ratios suggest a sympathetic flare event;
\item In the spectroscopic data a complex eruption including a CME was detected with falling-back and re-ejected material, with a maximal projected velocity of $\approx 675$\,\kms;
\item We estimate the mass of the ejecta higher that $10^{16}$\,g, which is comparable to the strongest solar CME masses;
\item The observed CME rate is much lower than expected from  extrapolation of the solar flare--CME relation to active stars, that could indicate that the solar--stellar analogy could not be applied to the cool dMe stars directly;
\item The spectral energy distribution does not suggest hot dust around the star that could be a result of high activity and strong stellar wind.
\end{itemize}

\begin{acknowledgements}
The authors would like to thank A. Mo\'or for the helpful discussions and the anonymous referee for the helpful comments that improved the paper significantly.
The authors acknowledge support from the Hungarian Research Grants 
OTKA K-109276, OTKA K-113117,
the Lend\"ulet-2009 and Lend\"ulet-2012 Program (LP2012-31)  of the Hungarian Academy of Sciences, and the ESA PECS Contract No. 4000110889/14/NL/NDe.
This research used the facilities of the Canadian Astronomy Data Centre operated by the National Research Council of Canada with the support of the Canadian Space Agency.

ML and PO  acknowledge  support  from  the  FWF project  P22950-N16.

This research used observations obtained at the Canada-France-Hawaii Telescope (CFHT) which is operated by the National Research Council of Canada, the Institut National des Sciences de l'Univers of the Centre National de la Recherche Scientique of France, and the University of Hawaii.

The Two Micron All Sky Survey (2MASS) is a joint project of the University 
of Massachusetts and the Infrared Processing and Analysis Center/California 
Institute of Technology, funded by the National Aeronautics and Space Administration 
and the National Science Foundation, USA. (\url{http://www.ipac.caltech.edu/2mass/})

This publication makes use of data products from the Wide-field Infrared Survey Explorer, which is a joint project of the University of California, Los Angeles, and the Jet Propulsion Laboratory/California Institute of Technology, funded by the National Aeronautics and Space Administration.

\end{acknowledgements}
\bibliographystyle{aa}
\bibliography{mn-jour,vida}

\appendix
\section{Detailed description of the complex CME event on HJD\,2453603}
\label{sect:cmedesc}
On HJD 2453603 a complex CME event was observed in the spectra (see dynamic spectra in Fig. \ref{fig:dynspecjd}, H, He lines in Fig. \ref{fig:cme} and intensity curves for the Balmer lines in Fig. \ref{fig:ha-lc}).
The chronology is as follows: blue wing enhancement \textnumero 1 (BWE1) starts at HJD 2453603.89 ($t_{1}$), lasts for nearly $\approx$ 30~min. and reaches a maximum projected velocity of $\approx -350$\,\kms. Also the red wing of H$\alpha$ is increased but not as significant as the H$\alpha$ blue wing. At the same time the H$\alpha$ profile starts to rise and marks the beginning of an impulsive flare phase. We detect a double-peaked H$\alpha$ line, with maxima at $-45$\,\kms{} and 15\,\kms, and also an additional narrower peak at $-65$\,\kms{} (see also Fig. \ref{fig:cme}). At the same time the H$\beta$ and H$\gamma$ lines rise dramatically and show a symmetrically broadened line profile, in contrast to the blue asymmetric broadened H$\alpha$ profile (BWE1). Moreover the He~{\sc i} line strength at 5876\,\AA{} also dramatically increases during the impulsive flare phase (see Figs. \ref{fig:cme} and \ref{fig:dynspecph-otherlines}). In the next spectrum $\approx15$ minutes later, the narrow peak cannot be seen any more, and the double peaked feature is also less prominent. In the next spectrum $\approx5$ minutes later the double peak almost disappears, and the continuum levels is also lowered. During the next one and a half hours the H$\alpha$ peak stays increased whereas the wings are not broadened anymore.

Then at HJD 2453603.966 ($t_{2}$), the blue wing of H$\alpha$ starts to rise again (BWE2) although the peak is still increased. The blue-wing asymmetry lasts for 50~minutes, and shows a maximum projected velocity of  $-350$\,\kms{} again. The red wing of H$\alpha$ increases much slower. Also the blue wings of H$\beta$ and H$\gamma$ increase and show projected velocities of $-275$\,\kms .

At HJD 2453604.001 ($t_{3}$), although the blue wing of H$\alpha$ did not reach its quiescent form, the H$\alpha$ blue wing starts to rapidly increase (BWE3) and forms a broad enhancement present for $\approx$45~minutes, showing a maximum projected velocity of $-675$\,\kms . And also the H$\beta$ and H$\gamma$ show similar blue wing enhancement, with maximum projected velocities of $-620$\,\kms .
After this turbulent events H$\alpha$, H$\beta$, and H$\gamma$ show red wing enhancements lasting for nearly 2 hours.

\section{Attempts on Doppler imaging}

We have tried to derive Doppler images from the CFHT spectra downloaded from the CFHT Science Archive with two codes: \texttt{TempMap} \citep{tempmapfirst} and \texttt{INVERS7PD} \citep{inverse7a,inverse7b}. In both cases, three consecutive spectra were averaged in order to increase signal-to-noise. The resulting smearing were always lower than the surface resolution of our reconstructions, i.e., $5^{\circ}$. Astrophysical parameters were adopted from \cite{Donati:2006:zdi}  and \cite{Morin:2008:zdi}. With \texttt{TempMap}, Kurucz atmospheric models \citep{kurucz} were used, while the \texttt{INVERS7PD} attempts were carried out with the MARCS models. Atomic parameters were obtained from the VALD database \citep{vald1, vald2}. 

The inversions were attempted on both Ca\,{\sc i}\,6439 and Fe\,{\sc i}\,6430. Other mapping lines in the region have not been used due to insufficient line strength or heavy blending with molecular bands. Ultimately, both inversion attempts proved unsuccessful. We suspect that the causes are the following:
\begin{itemize}
    \item Both the Kurucz and the MARCS models are insufficient in the domain of M dwarfs.
    \item It is possible, that due to the extremely strong chromospheric activity, the otherwise photospheric lines (mostly in case of the Ca\,{\sc i}\,6439) are burdened with emission cores.
    \item The region containing the lines used during the inversion is heavily affected by molecular bands.
\end{itemize}
Thus, synthesizing adequate model spectra and fitting them to the observed data proved virtually impossible. We conclude however, that an inversion attempt on TiO bands might be possible with data of higher S/N.

\end{document}